\documentclass[reprint,aps,prl,groupedaddress,10pt]{revtex4-2}

\usepackage{packages}
\usepackage{notations}

\begin{document}

\setlength{\abovedisplayskip}{4pt}
\setlength{\belowdisplayskip}{4pt}

\title{Inverse scattering solution of the weak noise theory of the Kardar-Parisi-Zhang equation with flat and Brownian initial conditions
}

\author{Alexandre Krajenbrink}
\email{alexandre.krajenbrink@sissa.it}
\affiliation{SISSA and INFN, via Bonomea 265, 34136 Trieste, Italy}
\author{Pierre Le Doussal}
\affiliation{Laboratoire de Physique de l'\'Ecole Normale Sup\'erieure, CNRS, ENS $\&$ PSL University, Sorbonne Universit\'e, Universit\'e de Paris, 75005 Paris, France}

\date{\today}

\begin{abstract}
We present the solution of the weak noise theory (WNT) for the Kardar-Parisi-Zhang equation in one dimension at short time 
for flat initial condition (IC). The non-linear hydrodynamic equations of the WNT are solved analytically through a connexion to the Zakharov-Shabat (ZS) system using its classical integrability. This approach is based on a recently developed Fredholm determinant framework previously applied to the droplet IC. The flat IC provides the case for a non-vanishing boundary condition of the ZS system and yields a richer solitonic structure comprising the appearance of multiple branches of the Lambert function. As a byproduct, we obtain the explicit solution of the WNT for the Brownian IC, which 
undergoes a dynamical phase transition. We elucidate its mechanism by showing that the related spontaneous breaking of the spatial symmetry arises
from the interplay between two solitons with different rapidities. 
\end{abstract}

\maketitle

Non-linear stochastic equations are a central tool in non-equilibrium physics \cite{vankampen}. 
They are often studied using optimal fluctuation theory and instanton methods
\cite{FreidlinWentzel,fogedby1998soliton,kleinert}.
This usually amounts to perform a saddle point evaluation on the action of the associated dynamical field theory
\cite{MSR,TauberBook}. 
In the favorable situations this approximation is controlled by a small parameter. This is often the case when describing
rare large fluctuations, i.e. large deviations
\cite{TouchetteReview2018,DerridaMFTReview2007}.
The resulting saddle point equations
are typically a set of coupled non-linear equations, which can only be solved in some special limits, or numerically.
It is rare that there are exact solutions, and even more remarkable when this set of equations is fully integrable. 

Recently we showed \cite{UsWNT2021} that the saddle point equations which describe the large deviations at short time for the Kardar-Parisi-Zhang (KPZ) 
stochastic growth equation in one space dimension, the so-called weak noise theory (WNT) \cite{Korshunov,Baruch,MeersonParabola,janas2016dynamical,meerson2017randomic,Meerson_Landau,Meerson_flatST,asida2019large,meerson2018large,smith2018finite,smith2019time,lin2020short}, can be solved exactly.
As noted in \cite{janas2016dynamical},
the basic system of non-linear equations
is the so-called $\{ P,Q \}$ system, a cousin of the non-linear Schrödinger equation. Using inverse scattering methods coupled to a recently developed Fredholm determinant framework \cite{krajenbrink2020painleve,bothner2021atlas} we showed how to
construct general solutions of this system, and obtained an explicit solution for the so-called droplet initial condition (IC) which is localized in space and
decays at infinity. In this paper we extend the method and present solutions in the case of initial conditions which are non-vanishing
at infinity. We first treat the case of the flat IC for the KPZ equation, from which, in a second stage, we obtain the 
solution for  the Brownian IC. 

The KPZ equation \cite{KPZ} describes the stochastic growth in time $\tau$ of the height field $h(y,\tau)$ of an interface,
here in one space dimension $y \in \mathbb{R}$
\be
\partial_\tau h(y,\tau) = \partial_y^2 h(y,\tau) + (\partial_y h(y,\tau))^2 + \sqrt{2} \eta(y,\tau) 
\label{eq:KPZ}
\ee
where $\eta(y,\tau)$ is a standard space time white noise, i.e. 
$\overline{\eta(y,\tau) \eta(y',\tau')}= \delta(\tau-\tau') \delta(y-y')$. In this work
we first focus on the solution of \eqref{eq:KPZ} with the flat IC 
\be
h(y,\tau=0)= 0
\ee
Because of the non-linear term in \eqref{eq:KPZ}, the growth at late times belongs to a different universality
class (the so-called KPZ class) than its simpler version, the Edwards-Wilkinson equation (without the non-linear term). 
Interestingly, this non-linearity has a profound effect already at short time, not for
the typical height fluctuations, which are Gaussian 
with Edwards-Wilkinson scaling $\delta h \sim \tau^{1/4}$, but for the
rare but much larger fluctuations $\delta h = \mathcal{O}(1)$.
For example, the probability $P(H,T)$ to observe the value of the field $h(0,T)=H$ 
at some time $\tau=T$,  takes, for $T \ll 1$
and $H=\mathcal{O}(1)$, the following large deviation form 
\be
P(H,T) \sim \exp( - \Phi(H)/\sqrt{T} ) 
\label{eq:HeightLargeDev}
\ee
The rate function $\Phi(H)$ was obtained analytically in a few cases where
Bethe ansatz solutions of the KPZ equation are available \cite{le2016exact,krajenbrink2017exact,krajenbrink2018large,KrajLedou2018,ProlhacKrajenbrink,krajenbrink2019beyond, doussal2019large}
and numerically in \cite{NumericsHartmann,hartmannprep}. 
In our previous work \cite{UsWNT2021} we showed how to obtain $\Phi(H)$ 
by solving exactly the weak noise theory equations. This is a completely different route, which until now
was limited to approximate or asymptotic solutions \cite{Korshunov,Baruch,MeersonParabola,janas2016dynamical,meerson2017randomic,Meerson_Landau,Meerson_flatST,asida2019large,meerson2018large,smith2018finite,smith2019time}. Being classically integrable, the $\{ P,Q \}$ system has an infinite number of conserved quantities, and
we showed that $\Phi(H)$ is obtained from one of them.  Solving the full equations gives much
more information beyond the rate function, since it determines the exact "optimal" KPZ height 
and noise space-time fields producing the rare fluctuations. Here we obtain these fields 
for the flat and Brownian IC as well as the rate functions. We follow the
same outline as in \cite{UsWNT2021} and indicate how the present case
differs in crucial ways. 

Let us recall how the $\{ P,Q \}$ system arises. It is more convenient to 
work with the exponential field $Z=e^{h(y,\tau)}$.
It is also equal to the partition sum of a directed polymer $x(\tau)$ at equilibrium in a random potential $\eta(x(\tau),\tau)$ (the KPZ noise)
in dimension $d=1+1$. The equivalence of the two problems is quite convenient, e.g. for
numerical simulations \cite{NumericsHartmann, hartmannprep}. 
We have introduced the rescaled time and space variables as $t=\tau/T$, $x=y/\sqrt{T}$,
where $T$, the observation time, is fixed \cite{footnoteScalingXT}. The field $Z(x,t)$, expressed in these coordinates, satisfies the (rescaled) stochastic heat equation (SHE) in the Ito sense
\be
\partial_t Z(x,t) = \partial_x^2 Z(x,t) + \sqrt{2} T^{1/4} \tilde \eta(x,t) Z(x,t) 
\label{eq:SHE}
\ee
Here $\tilde \eta(x,t)$ is another standard space time white noise. This equation 
is now studied for $t \in [0,1]$. The noise amplitude being now $\mathcal{O}(T^{1/4})$, 
a short observation time $T \ll 1$ corresponds to a weak noise.
As in \cite{le2016exact,krajenbrink2017exact,krajenbrink2018large,KrajLedou2018,ProlhacKrajenbrink,krajenbrink2019beyond,doussal2019large}
and in \cite{UsWNT2021}, it
is convenient to study the following generating function
which admits a large deviation principle at short time $T \ll 1$, with $z \geq 0$ 
\be
\overline{ \exp\big( -z e^H /\sqrt{T}  \big)  }  \sim \exp\big( - \Psi(z)/\sqrt{T} \big) 
\label{eq:ZLargeDev}
\ee 
Inserting \eqref{eq:HeightLargeDev} into the expectation value over $P(H,T)$ in the l.h.s., we see that for $T \ll 1$, 
$\Psi(z)$ and $\Phi(H)$ are related through a Legendre transform
\be  \label{Legendre} 
\Psi(z)=\min_H ( z e^H + \Phi(H) ) 
\ee
As detailed in \cite{UsWNT2021}, in the short time limit $T \ll 1$
the expectation value \eqref{eq:ZLargeDev} over the 
the stochastic dynamics \eqref{eq:SHE} can be obtained from saddle point equations,
which take the form of the $\{ P,Q\}_g$ system
\be
\begin{split}\label{eq:PQsystem}
 \partial_t Q =& \partial_x^2 Q + 2 g P Q^2  \\
 - \partial_t P =& \partial_x^2 P + 2 g P^2 Q  
 \end{split}
\ee
These equations for $P(x,t)$, $Q(x,t)$ must be solved for $x \in \mathbb{R}$ and 
$t \in [0,1]$ with mixed boundary conditions, which
for the flat IC read
\be \label{init}
Q(x,0) = 1  , \quad P(x,1)=\delta(x)
\ee
and the coupling set to $g=-z$. The new feature, as compared
to \cite{UsWNT2021} is that $Q(x,t) \to 1$ for $x \to \pm \infty$.
The function $P$ however, as well as the product $P Q$, still decay at
infinity. The solution of \eqref{eq:PQsystem}, \eqref{init} determines the
optimal height via $Z_{\rm opt}(x,t)=e^{h_{\rm opt}(y,\tau)}=Q(x,t)$ while
the optimal noise is $\tilde \eta_{\rm opt}(x,t) = P(x,t)Q(x,t)$. 
As in \cite{UsWNT2021} we will calculate from the solution the 
value $C_1(g)$ of the first conserved quantity, $C_1 = g \int_\R\rmd x P Q$,
and from $C_1(g)$ obtain the rate function $\Psi(z)$. Indeed $C_1$
being time independent, at $t=1$ one has $C_1(g)=g Q(0,1) = g e^H$. 
On the other hand, differentiating the Legendre transform in \eqref{Legendre}
w.r.t. $z$ gives $\Psi'(z)=e^H$. Since $g=-z$ this gives $C_1(-z) = - z \Psi'(z)$,
from which we obtain $\Psi(z)$ by integration, and, in a second stage, $\Phi(H)$ by
Legendre inversion of \eqref{Legendre}.

As in \cite{UsWNT2021}, to solve the non linear system \eqref{eq:PQsystem}, \eqref{init} one proceeds
in two stages: the direct and the inverse scattering problems. First one studies an auxiliary scattering problem \cite{ZS}, in which the
scattering amplitudes obey a linear time evolution, and exhibit a very simple time dependence. 
In a second stage, from these scattering amplitudes, one constructs the solution of \eqref{eq:PQsystem}, \eqref{init}. 
The $\{P,Q\}_g$ system belongs to the AKNS class 
\cite{AblowitzKaup1974}, for which there exists a Lax pair, i.e. a pair of linear
differential equations whose compatibility conditions are equivalent to \eqref{eq:PQsystem}.
Here the system reads $\partial_x \vec v= U_1 \vec v$, $\partial_t \vec v= U_2 \vec v$
where $\Vec{v}=(v_1,v_2)^\intercal$ is a two component vector (depending on space, time and spectral variables $x,t,k$) 
where
\begin{equation}
U_1=
\begin{pmatrix}
-\I k/2  & - g P(x,t)\\  Q(x,t) & \I k/2 
\end{pmatrix} \quad , \quad 
U_2= 
\begin{pmatrix}
{\sf A} & {\sf B}\\
{\sf C} & -{\sf A}
\end{pmatrix}
\label{eq:LaxPairU}
\end{equation}
where ${\sf A}= k^2/2- g P Q$, ${\sf B}=g (\partial_x - \I k) P$,
${\sf C}= (\partial_x +  \I k) Q$. One can check that the compatibility condition, $\p_t U_1-\p_x U_2 +[U_1, U_2]=0$,
recovers the system \eqref{eq:PQsystem}. In particular, the Lax pair implies the existence of an infinite number of conserved quantities. The new feature as compared to \cite{UsWNT2021} is that for
the flat IC we have $Q(\pm \infty,t)=c$ (we set $c=1$ later) hence the eigenvectors at $x=\pm \infty$ of the matrix $U_1$ are now
$(1,c/(-\I k))^\intercal$ and $(0,1)^\intercal$ with eigenvalues $- \I k/2$ and $\I k/2$ respectively.
We define two linearly independent pairs of solutions of the $x$ member of the Lax pair as 
$\Vec{v}=e^{ k^2 t/2} {\phi}$ with $ {\phi}=(\phi_1,\phi_2)^\intercal$ and $\Vec{v}=e^{- k^2 t/2} {\bar{\phi}}$
with $ {\bar \phi}=(\bar \phi_1, \bar \phi_2)^\intercal$
for the first pair, and $\phi, \bar \phi$ replaced by $\psi, \bar \psi$ for the second pair. The
pair $\phi, \bar \phi$ is such that 
at $x \to -\infty$, $\phi \simeq e^{-\I k x/2} (1,c/(-\I k))^\intercal$ and $\bar \phi \simeq e^{\I k x/2} (0,-1)^\intercal$.
The pair $\psi, \bar \psi$ is such that 
at $x \to +\infty$, $\psi \simeq e^{\I k x/2} (0,1)^\intercal$ and $\bar \psi \simeq e^{- \I k x/2} (1,c/(- \I k))^\intercal$.
The linear relation between the two independent pairs of solutions defines the four scattering amplitudes
\begin{equation}
\begin{split}
\phi(x,t,k)&=a(k,t)\bar{\psi}(x,t,k)+b(k,t)\psi(x,t,k)\\
\bar{\phi}(x,t,k)&= \tilde{b}(k,t)\bar{\psi}(x,t,k)-\tilde{a}(k,t)\psi(x,k,t)
\end{split}
\end{equation}
Equivalently, this implies the following asymptotics for $\phi, \bar \phi$ at $x=+\infty$ 
\bea \label{eq:plusinfinity} 
&& \phi \underset{x \to +\infty}{\simeq}
\begin{pmatrix}
a(k,t)e^{-\frac{\I  kx}{2}}\\b(k,t)e^{\frac{\I  kx}{2}} + \frac{c}{- \I k} a(k,t)e^{-\frac{\I  kx}{2}}
\end{pmatrix}  \\
&& 
\bar{\phi} \underset{x \to +\infty}{\simeq}
\begin{pmatrix}
\tilde{b}(k,t)e^{-\frac{\I  kx}{2}}\\ -\tilde{a}(k,t)e^{\frac{\I  kx}{2}} + \frac{c}{- \I k}
\tilde{b}(k,t)e^{-\frac{\I  kx}{2}}
\end{pmatrix} \nn 
%\end{cases}
%\end{split}
\eea
Plugging this form into the $\partial_t$ equation of the Lax pair at $x \to +\infty$,
one finds %that these amplitude have a 
a very simple time dependence, %i.e.
$a(k,t)=a(k)$ and $b(k,t)=b(k)e^{-k^2 t}$, 
$\tilde a(k,t)=\tilde a(k)$ and $\tilde b(k,t)=\tilde b(k)e^{k^2 t}$.
The Wronskian $W=\phi_1 \bar{\phi_2}-\phi_2 \bar{\phi_1}$ is
space and time independent since
$\p_x W=\Tr(U_1) W=0$ and $\p_t W=\Tr(U_2) W=0$. It is
$W=-1$ at $x=-\infty$ and evaluating it using \eqref{eq:plusinfinity}
at $x=+\infty$ leads to the relation
\begin{equation}
a(k)\tilde{a}(k)+b(k)\tilde{b}(k)= 1
\end{equation}
as in the case $c=0$. 

Let us now make use of the boundary data in \eqref{init}, and characterize the scattering amplitudes. Integrating the
$\partial_x$ equation of the Lax pair at $t=1$ for $\bar{\phi}$ 
using \eqref{init} allows to obtain \cite{SM} $\tilde{b}(k)=g e^{-k^2}$,
together with some relations between $a(k)$ and $\tilde a(k)$ and
$Q(x,1)$ (which is yet unknown). Using that $Q(x,0)=c=1$, is even in $x$
then $\tilde{a}(k)=a(-k)=a^*(k^*)$
and $b(k)$ is real and even. This leads to the form
\begin{equation}
a(k)=e^{- \I \varphi(k)} \sqrt{1-g b(k) e^{-k^2}}
\end{equation}
where we still have two unknown functions, a phase $\varphi(k)$, which is odd $\varphi(k)=-\varphi(-k)$, and $b(k)$.

The form for the amplitudes obtained at this stage are still quite similar to the general solution for 
decaying IC (i.e. of the droplet type) obtained in \cite{UsWNT2021}. For the droplet IC we obtained $b(k)=1$.
Here we obtain $b(k)$ for the flat IC as follows. Let us return to the $\partial_x$ equation at $t=0$ using 
that $Q(x,0)=c$. It reads
\begin{equation} \label{eq:barphi12n0} 
\partial_x {\phi}_1=-\I \frac{k}{2} {\phi}_1- g P(x,0) {\phi}_2 \quad , \quad 
\partial_x {\phi}_2=\I \frac{k}{2} {\phi}_2+ c {\phi}_1\\
\end{equation}
Eliminating $\phi_1$ we obtain
\bea \label{eq:schrod} 
\partial^2_x {\phi}_2 + (c g P(x,0) + \frac{k^2}{4} ) \phi_2 = 0
\eea 
Unlike the general case, it is a Schrödinger equation with a {\it real} potential. Hence if
$\phi_2$ is solution, so is $\phi_2^*$. Note that $\bar \phi_2$
satisfies also \eqref{eq:barphi12n0} and \eqref{eq:schrod}.
For $x \to -\infty$, from the aforementioned 
asymptotics, one has $\phi_2^* = \frac{c}{- \I k} \bar \phi_2$. Hence
the same relation should hold for any $x$, including $x \to +\infty$. 
From \eqref{eq:plusinfinity} one then obtains $a^*(k^*)=\tilde a(k)$,
which we already knew, and 
\be \label{b} 
b(k)= - \frac{c^2}{k^2} \tilde b(k) = -  \frac{g}{k^2} e^{-k^2} 
\ee
where we set $c=1$ in the last identity.

It remains to obtain $\varphi(k)$. Here we will rely on \cite{UsWNT2021} where for 
a general $b(k)$ we obtained 
\be  \label{phi1} 
\varphi(k)= \dashint_\R \frac{\rmd q}{2\pi} \, \frac{k }{q^2-k^2}\log(1- g b(q) e^{-q^2} )
\ee 
The proof presented there was based on a random walk representation which assumes that $b(q)$ has a proper
inverse Fourier transform. It thus cannot be readily applied here. We believe that this is a technical
issue (which maybe can be resolved using proper regularizations) and we will here {\it conjecture} that 
\eqref{phi1} extends to the present case. This conjecture will be abundantly confirmed by the
results below.

Having determined the scattering amplitudes we now follow \cite{UsWNT2021} to perform the inverse-scattering transform,
and obtain the solution of the $\{P,Q\}_g$ system \eqref{eq:PQsystem} for the flat IC \eqref{init} as 
\bea \label{soluQP} 
&& Q(x,t)= \bra{ \delta } {\cal A}_{xt} (I + g {\cal B}_{xt} {\cal A}_{xt})^{-1} \ket{\delta} \\
&& P(x,t)= \bra{ \delta } {\cal B}_{xt} (I + g {\cal A}_{xt} {\cal B}_{xt})^{-1} \ket{\delta} \nonumber
\eea
where $\ket{\delta}$ is the vector with component $\delta(v)$ so that $\bra{\delta} {\cal O}\ket{\delta}=\mathcal{O}(0,0)$ for
any operator $\mathcal{O}$. Here ${\cal A}_{xt}$, ${\cal B}_{xt}$ are 
two linear operators from $\mathbb{L}^2(\mathbb{R}^+)$ to $\mathbb{L}^2(\mathbb{R}^+)$ with respective kernels
\be \label{kernels} 
{\cal A}_{xt}(v,v')= A_t(x+v+v') ~,~ {\cal B}_{xt}(v,v')= B_t(x+v+v')
\ee
where the two functions $A_t(x)$ and $B_t(x)$ are the Fourier transform of the time-dependent reflection coefficients and obey the heat equation (and, respectively, its time reverse) and are given for $g<0$ by
\bea 
 && A_t(x)=- g \int_\R \frac{\rmd k}{2\pi} \frac{e^{\I k x-k^2(1+ t)+\I \varphi(k)} }{k^2\sqrt{1+ g^2 k^{-2} e^{-2 k^2}}}  + \frac{1}{2} 
 \label{AGeneral} \\
 && B_t(x)=\int_\R \frac{\rmd k}{2\pi} \frac{e^{-\I k x-k^2(1- t)-\I \varphi(k)}}{\sqrt{1+ g^2 k^{-2} e^{-2 k^2}}} \label{BGeneral}
\eea
Here the phase reads
\be \label{soluphase}
\varphi(k)= \dashint_\R \frac{\rmd q}{2\pi} \, \frac{k }{q^2-k^2}\log(1+g^2 q^{-2} e^{-2 q^2} )
\ee

\begin{figure*}[t!]
    \centering
    \includegraphics[scale=0.7]{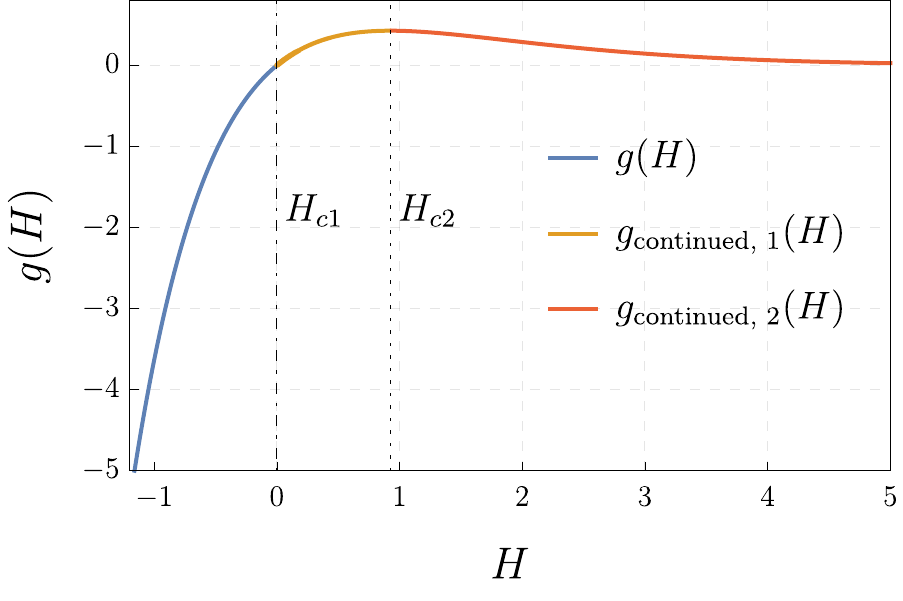}
    \includegraphics[scale=0.2, trim=0 -5cm 0 0, clip]{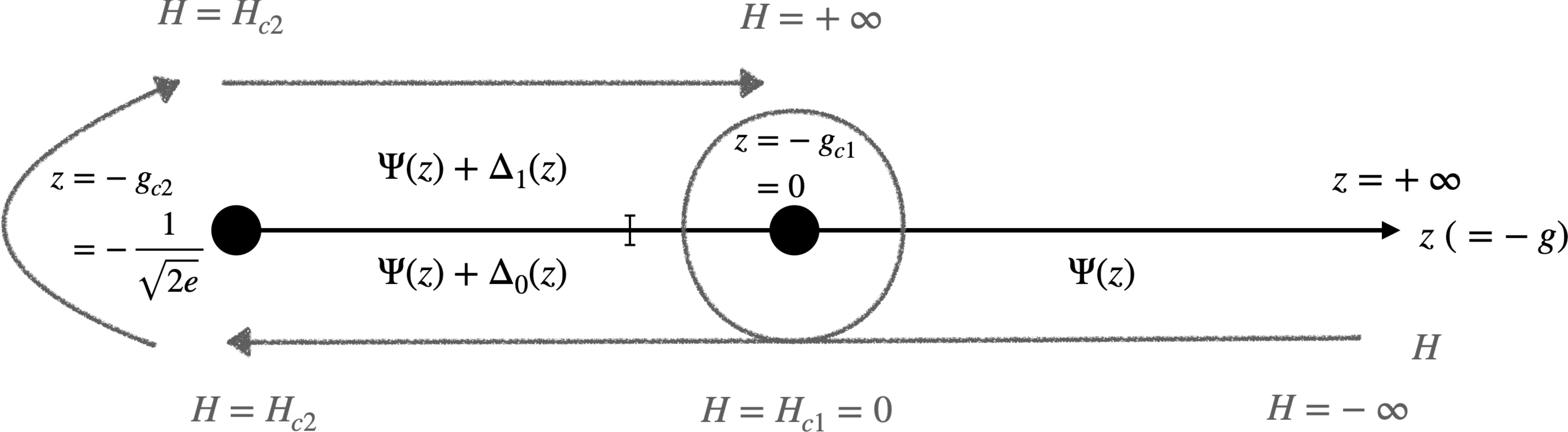}
    \caption{\textbf{Left.} Plot of the coupling $g=g(H)$ of the $\{P, Q \}$ system \eqref{eq:PQsystem}, 
    \eqref{init} to be used to obtain $\Phi(H)$ for a given $H$. It is obtained from $H=\log \Psi'(z=-g)$.
    The fields $H_{c1}=0$ and $H_{c2}= 0.926581$ correspond to the limits of the three branches of
    solutions discussed in the text (with $g_{c1}=0$ and $g_{c2}=1/\sqrt{2 e}$).
      \textbf{Right.} Schematic plot of the branches for $\Psi(z)$ as $z=-g$ is varied, and the corresponding
      ranges of values for $H$. For $H < H_{c_1}=0$ one uses $\Psi(z)=\Psi_0(z)$ given in Eq.~\eqref{eq:psi0_z}.
      At $H = H_{c_1}=0$, one needs to turn around the branching point of $\Psi_0(z)$ at $z=0$,
      and change the Riemann sheet. This leads to the continuation $\Psi(z)=\Psi_0(z)+\Delta_0(z)$ 
      which, using Eqs.~\eqref{inversion}, determines $\Phi(H)$ for all $H_{c1}=0<H<H_{c2}$ by decreasing $z$ from $0$ to $-g_{c2}$.
      A second continuation $\Psi(z)=\Psi_0(z)+\Delta_1(z)$ is obtained similarly by rotating around $z=-g_{c2}$.
      which, using Eqs.~\eqref{inversion}, determines $\Phi(H)$ for all $H_{c2}<H$ by increasing $z$ from $-g_{c2}$ back to $0$.}
    \label{fig:branches_g_versus_h}
\end{figure*}

We used \cite[Eq.~(11)]{UsWNT2021} inserting the scattering data obtained above. Note however the additional constant $1/2$ in \eqref{AGeneral}. 
Indeed, since the product ${\cal A}_{xt} {\cal B}_{xt}$ vanishes for $x \to +\infty$, 
one must have $Q(x,t) \simeq A_t(x)$ for $x \to +\infty$. Since $Q(\pm \infty,x)=1$ we must have $\lim_{x \to +\infty} A_t(x) = 1$.
We have checked that this is indeed the case from \eqref{AGeneral}, the $1/2$ constant being crucial. Its origin can
be traced to the pole in the integrand of \eqref{AGeneral}, following the general scheme in \cite{ZS}. 
The functions $\varphi(k)$ and $A_x(t),B_x(t)$ are plotted in \cite{SM} for various values of $t,g$.
Note that $\varphi(0^\pm)=\mp \frac{\pi}{2}$ for any $g\neq 0$ so that the integrand in $A_t$ behaves as $- {\rm sgn}(g)/(\I k)=\frac{1}{\I k}$ 
since \eqref{AGeneral},\eqref{BGeneral} are valid only for $g<0$, at small $k$.
We further note the unexpected relation $A_t''(x) = g B_{-t}(x)$. 

We can now examine the conserved quantities $C_n$ and obtain $\Psi(z)$ from $C_1$.
The $C_n$ for the $\{ P,Q\}_g$ system were obtained in \cite{UsWNT2021}, with
$C_1= g \int_\R\rmd x P Q$, $C_2=g \int_\R\rmd x P \p_x Q$, $C_3=g (\int_\R\rmd x P \p_x^2Q + g P^2 Q^2)$
and so on. Since the product $P Q$ still vanishes at infinity, these remain valid in the present case. 
As before, the values $C_n(g)$ of these conserved charges can be extracted \cite{SM} by expanding $- \I \varphi(k) = \sum_{n \geq 1} \frac{C_n(g)}{(\I k)^n}$ in powers of $1/k$ 
in \eqref{soluphase}. This leads to 
$C_{2m +1}(g) = (-1)^{m-1} \int_\R\frac{\rmd q}{2 \pi} q^{2m} \log(1 + g^2 q^{-2} e^{-2 q^2} )$. Since $- z \Psi'(z) = C_1(-z)$, with $g=-z$, we obtain $ - z \Psi'(z) 
= \int_\R \frac{\rmd q}{2 \pi} {\rm Li}_1(- \frac{z^2}{q^2}  e^{-2q^2})$, where 
${\rm Li}_1(y)=- \log(1-y)$. Using the relation between polylogarithm functions, $z \partial_z {\rm Li}_n = {\rm Li}_{n-1}$, 
we obtain upon integration, our final result for the flat IC
\be \label{eq:psi0_z}
\Psi(z) = \Psi_0(z) := - \int_\R \frac{\rmd q}{4 \pi} {\rm Li}_2(- \frac{z^2}{q^2}  e^{-2q^2}) 
\ee
%Note that $C_{2 n+1} = \int_\R \rmd x P(x,0)^n$. 
Taking a derivative of \eqref{Legendre} one obtains the rate function
$\Phi(H)$ in a parametric form 
\be \label{inversion} 
e^H = \Psi'(z)  \quad , \quad \Phi(H) =  \Psi(z) - z \Psi'(z)
\ee
As in \cite{UsWNT2021} this is valid only for $z>0$ (i.e. $g<0$) since the r.h.s. in \eqref{eq:ZLargeDev} diverges 
for $z<0$. Since $\Phi'(H) = - z e^{H}$, the range $z>0$ corresponds to $H$ in $(-\infty,0]$, where
$H=0$ is the most probable value of $H$ defined by $\Phi'(H)=0$. Thus up to now
we have solved the case $g<0$, i.e. $z>0$ which corresponds to the left side of $P(H,t)$
and to the {\it main branch} for $\Psi(z)$.

To obtain the right side $H>0$ we proceed as in \cite{UsWNT2021}. 
Equations~\eqref{eq:PQsystem} also hold for any $H>0$, corresponding 
to the attractive regime $g>0$ of the $\{ P,Q\}_g$ system. 
Indeed, $\Psi(z)$ can be analytically continued to $z<0$, allowing to 
determine $\Phi(H)$ for any $H$. By contrast with the droplet IC, the
flat IC requires a continuation in two steps. Since $\Psi_0(z)$
has a branch cut on the negative real axis, for 
$H \in [0,H_{c2}]$, with $H_{c2} = 0.926581$, see \cite[Eq.~(S39)]{SM}, 
a first continuation is needed, with $\Psi(z)=\Psi_0(z)+ \Delta_0(z)$ (second branch),
where $\Delta_0(z)$ is obtained from the cut of $\Psi_0(z)$. In that branch, $g=-z$ increases from $0$ to $g_{c2}$, with $g_{c2}= 1/\sqrt{2 e} = 0.428882$. This is further explained in Fig.~\ref{fig:branches_g_versus_h}. 

For $H \in [H_{c2},+\infty]$, a third branch is required, $\Psi(z)=\Psi_0(z)+ \Delta_1(z)$ and $g=-z$ now decreases from $g_{c2}$ back to $0$, see Fig.~\ref{fig:branches_g_versus_h}. 
These continuations correspond to two branches of solutions of the $\{ P,Q\}_g$ system for $0<g \leq g_{c2}$.
As in \cite{UsWNT2021} these branches have a very nice physical origin, and one
one finds that the second branch corresponds to the spontaneous generation of a soliton while the third one is interpreted as a modification of the rapidity of the soliton.
In all branches, the rate function $\Phi(H)$ is obtained from \eqref{inversion} by
inserting the corresponding result for $\Psi(z)$, i.e. $\Psi_0$ for the main branch, $\Psi_0+\Delta_0$ and $\Psi_0+\Delta_1$
for the second and third branches.

Technically, the second branch arises from the fact that, for $g>0$, the logarithm inside $\varphi(k)$ has a cut for the integration variable in Eq.~\eqref{soluphase} located at  $q=\pm \I \kappa_0$ with
\be 
\label{eq:rapidity_root}
\kappa_0^2 e^{-2 \kappa_0^2} = 1/g^2  \quad , \quad  \kappa_0^2 = -\frac{1}{2} W_0(- 2 g^2) 
\ee 
where $W_0$ is the Lambert function \cite{corless1996lambertw} and $\kappa_0$ is  the positive root of \eqref{eq:rapidity_root}. This cut exists only if $0< g \leq g_{c2}$. The third branch arises from the continuation of the Lambert function $W_0$ to $W_{-1}$ so that the position of the cut is located at $q=\pm \I \kappa_1$ with $\kappa_{1}^2 = -\frac{1}{2} W_{-1}(- 2 g^2)$, see \cite{SM}. The contribution of the cuts give rise to a pole in the integrand of $A_t$ (resp. $B_t$) in the upper (resp. lower) half plane which according to the general construction of \cite{ZS} simply generates solitons.

Practically, the cuts of the phase $\varphi$ modify the expression of $A_t$ and $B_t$ by adding rational factors providing poles whose residues generate the solitons, see \cite[Supp Mat - Section S-K.]{UsWNT2021}. For the second branch, $0<g <g_{c2}$ and $0<H<H_{c2}$, one finds
\begin{widetext}
\bea \label{ABGeneralContinued} 
 && A_t(x)=- g \int_\R \frac{\rmd k}{2\pi} \frac{e^{\I k x-k^2(1+ t)+\I \varphi(k)} }{k^2\sqrt{1+ g^2 k^{-2} e^{-2 k^2}}}\frac{k+\I \kappa_0}{k-\I \kappa_0} + \frac{1}{2} +\frac{2g}{\kappa_0}e^{-\kappa_0 x +\kappa_0^2 (1+t)+\I \varphi(\I \kappa_0)}\\
 && B_t(x)=\int_\R \frac{\rmd k}{2\pi} \frac{e^{-\I k x-k^2(1- t)-\I \varphi(k)}}{\sqrt{1+ g^2 k^{-2} e^{-2 k^2}}} \frac{k-\I \kappa_0}{k+\I \kappa_0}+2\kappa_0e^{-\kappa_0 x +\kappa_0^2 (1-t)-\I \varphi(-\I \kappa_0)} \nn
\eea
\end{widetext}
where $\varphi(k)$ is given in Eq.~\eqref{soluphase}. The cuts also modify the conserved quantities by adding a solitonic contribution $\Delta C_n(g) = \frac{2}{n} \kappa_0^n$ for $n$ odd
and zero even charges \cite{UsWNT2021}. Integrating $-z \Delta_0'(z)= \Delta C_1(g=-z)$ one finds
\begin{equation} \label{De0} 
    \Delta_0(z)=\frac{\sqrt{2}}{3}[-W_0(-2z^2)]^{3/2}-\sqrt{2}[-W_0(-2z^2)]^{1/2}
\end{equation}

The third branch, $0<g <g_{c2}$ and $H>H_{c2}$ is obtained by the minimal replacement of $\kappa_0$ by $\kappa_1$ in both functions $A_t(x)$ and $B_t(x)$
in \eqref{ABGeneralContinued}. This leads again to $\Delta C_1(g) = 2 \kappa_1$ and, by integration,  
to $\Delta_1(z)$ given by the same equation as \eqref{De0} with $W_0 \to W_{-1}$. As in \cite{UsWNT2021} 
the solitonic part dominates the large deviations for $H \to +\infty$.

\begin{figure*}[t!]
    \centering
    \includegraphics[scale=0.5]{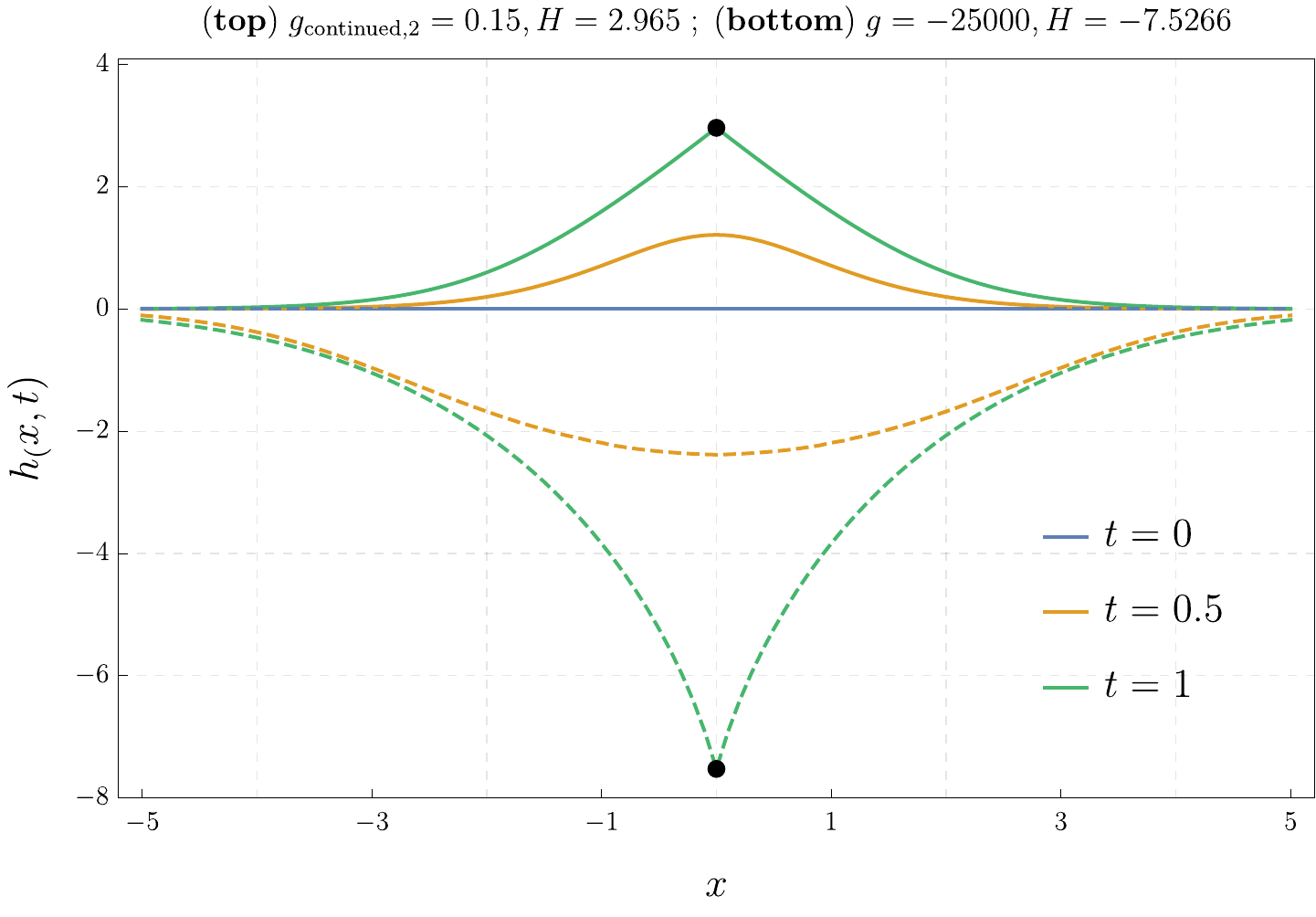}
    \hfill
    \includegraphics[scale=0.5]{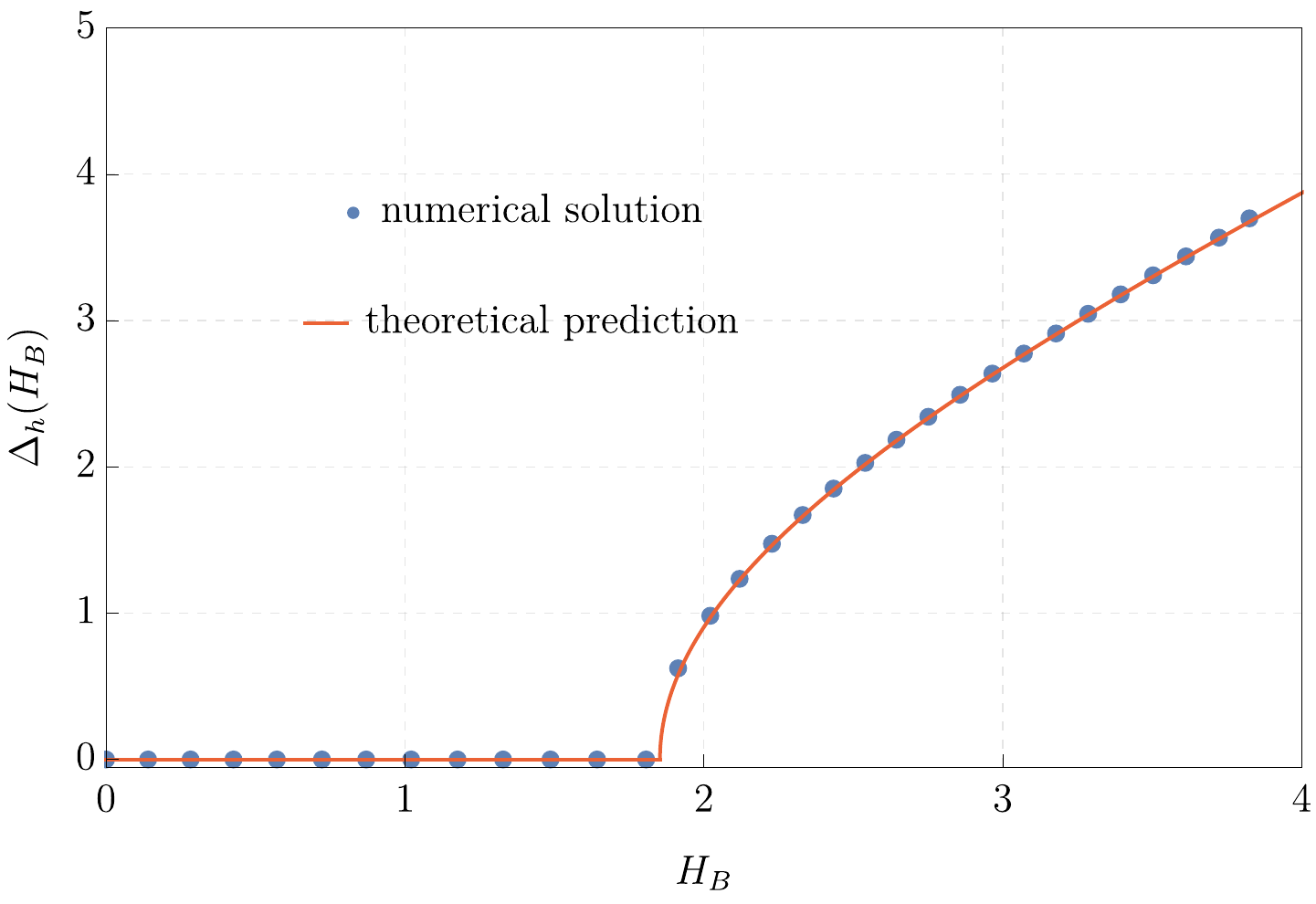}
    \caption{(\textbf{Left}) The optimal height $h_{\rm opt}(x,t)$ for flat initial conditions plotted for various times $t$ and for two values of $H$
    indicated by the black dots (one in the main branch $H=-7.5266$ - full line - and the other in the third branch $H=2.965$ - dashed line). (\textbf{Right})
    Plot of the order parameter $\Delta_h$ of the parity breaking transition for the Brownian IC as a function of 
    $H_B$ predicted here in \eqref{OP}, as compared to $-\Delta/2$, 
    where $\Delta$ is defined and obtained numerically in \cite{janas2016dynamical}. Courtesy of B.~Meerson for the data of the numerical solution of the WNT equations. }
    \label{fig:hxt}
\end{figure*}

From the above exact solutions for $A_t(x)$ and $B_t(x)$ we obtain the solutions to 
the $\{P,Q\}_g$ system through the Fredholm operator inversion formula \eqref{soluQP} for various
values of $H$ and $g$. We use the numerical method in \cite[Section S-L]{UsWNT2021}.
We have performed several numerical checks of some highly non-trivial consequences of the formulae,
which validate our conjecture: \textit{(i)} the functions $P,Q$ are even in $x$, \textit{(ii)}
$Q(x,t=1)=A_1(|x|)$, \textit{(iii)} $Q(0,t=1)=e^H=\Psi'(-g)$, and \textit{(iv)} $Q(\pm \infty,t)=1$.
The results for the optimal height $h_{\rm opt}(x,t)=\log Q(x,t)$ are plotted in
Fig.~\ref{fig:hxt}. 

The above results provide the first direct analytical derivation of $\Phi(H)$ for the flat IC.
Note that they are in agreement with those of Ref.~\cite{Meerson_flatST} which were cleverly inferred,  using various symmetries of the weak noise theory together with the known rate function of the Brownian initial condition calculated from the Bethe ansatz in \cite{krajenbrink2017exact}.

Conversely, starting from the flat IC, a remarkable byproduct of our results is the solution of the WNT for the KPZ equation with the Brownian (i.e stationary) IC.
It is defined as the solution of \eqref{eq:KPZ}  with $h(y,0)=W(y)$ where $W(y)$ is a two-sided standard Brownian
motion with zero drift with $W(0)=0$. This corresponds to \eqref{eq:SHE} with initial condition
$Z(x,0)=e^{T^{1/4} W(x)}$. We are interested in the probability $P(H_B,\tau)$ that $h(0,\tau)=H_B$,
which behaves at small $t$ as $P(H_B,T) \sim \exp(- \Phi(H_B)/\sqrt{T})$. 
To obtain the solution in that case we first notice that %A remarkable feature of 
our solution for the flat IC for $P(x,t)$, $Q(x,t)$ defined in \eqref{soluQP} is also well defined in the extended interval $t \in [-1,1]$, since
the equations \eqref{AGeneral} and \eqref{BGeneral} are also well defined in this interval. 
Let us now define the functions $P_B$ and $Q_B$ for $t \in [0,1]$ as
\be
\begin{split}
    \label{PQB_flat} 
&Q_B(x, t)= e^{\frac{H_B}{2} } Q(\sqrt{2} x, 2 t - 1) \\
&P_B(x, t) = \sqrt{2} P(\sqrt{2} x, 2 t - 1)
\end{split}
\ee
One can check that $P_B,Q_B$ satisfy the $\{P,Q\}_{g_B}$ system \eqref{eq:PQsystem} with coupling constant $g_B=\sqrt{2} g e^{-H_B/2}$  
and $Q_B(0,1)=e^{H_B}$ with $H_B=2 H$. As we show in \cite{SM}, they obey the boundary conditions
\textit{(i)} $Q_B(0,0)=1$, \textit{(ii)} $P_B(x,1)=\delta(x)$, \textit{(iii)}
\be \label{boundaryB} 
g_B P_B(x,0) Q_B(x,0) + \partial_x^2 \log Q_B(x,0) = g_B e^{H_B} \delta(x) 
\ee 
as well as \textit{(iv)} $Q_B(0,1)=e^{H_B}$. As shown in \cite{janas2016dynamical} the boundary conditions \textit{(i)}-\textit{(iv)}
for the $\{P,Q\}_{g_B}$ system are the one corresponding to the Brownian IC, hence $P_B,Q_B$ constructed as above provide the
solution of the WNT in that case. 

We have thus obtained through \eqref{PQB_flat} the solution for the Brownian IC in terms
of our solution $P,Q$ for the flat IC (extended in $t \in [-1,1]$). Let us discuss now what happens
for the different branches as $H_B=2 H$ is varied. In the main branch, $H \leq 0$, the function
$A_t,B_t$ are given by \eqref{AGeneral} and \eqref{BGeneral}. A consequence (see \cite{SM}) is that $\Phi(H)= \frac{1}{2 \sqrt{2}} \Phi_B(2 H)$
for $H \leq 0$ (in fact for $H \leq H_{c_2}$ see below). 

The discussion of the other branches is a bit more involved in
the Brownian case. For the second branch $0 < H = \frac{H_B}{2} <H_{c2} $ the construction is exactly the same as for the flat IC, i.e. one uses
\eqref{PQB_flat} and in $P,Q$ 
one chooses the continuations for $A_t,B_t$ given in \eqref{ABGeneralContinued} which includes the solitonic part 
with rapidity $\kappa_0$. For the third branch $H = \frac{H_B}{2} > H_{c2}$ it is in principle allowed to proceed to the change $\kappa_0 \to \kappa_1$ solely in one of the functions $A_t$ or $B_t$: hence there exist two additional distinct asymmetric solutions that we did not consider for the flat IC. In that case
the solutions $P,Q$ will not be even in $x$, providing a mechanism for a spontaneous breaking of the symmetry $x \to -x$. This was forbidden for the
flat IC, which is why one must choose $\kappa_0 \to \kappa_1$ in both $A_t,B_t$, leading to an even solution. 
For the Brownian IC however, it was shown \cite{krajenbrink2017exact} that the large deviation function $\Phi_B(H_B)$ has a second-order 
phase transition at precisely this value $H=H_{c2}$. The solution obtained here provides a mechanism for this transition.
As was observed in \cite{janas2016dynamical} this phase transition is indeed accompanied by a spontaneous symmetry
breaking of the spatial parity in the $\{P_B,Q_B\}_{g_B}$ solution, although no analytical results were obtained there for $H_B \approx 2 H_{c_2}$.
Hence for the Brownian IC we claim that there are two equivalent solutions, denoted $\pm$, related by parity,
i.e. $P_B^-(x,t)=P_B^+(-x,t)$, $Q_B^-(x,t)=Q_B^+(-x,t)$, and
which are obtained by replacing solely one $\kappa_0$ into a $\kappa_1$ inside either $A_t$ ($+$) or $B_t$ ($-$)
and using \eqref{PQB_flat}.

This is further understood from the solitonic contributions to the conserved quantities of the $\{P,Q\}_g$ system
given for all $n$ in this case as \cite[Supp Mat - Eq.~(S59)]{UsWNT2021}
\be  \label{cons} 
\Delta C^{\pm }_n = \pm \frac{1}{n} (\kappa_0^n - (-\kappa_{1})^n) 
\ee  
For $n=1$ this implies that the corresponding value of $\Psi(z)$ for this asymmetric solution is
$\Psi(z)=\Psi_0(z) + \frac{\Delta_0(z)+\Delta_1(z)}{2}$, which gives $g(H)$ in that branch from
$\Psi'(z=-g)=e^{H}$, in agreement with \cite{krajenbrink2017exact}, see \cite{SM}. Note that now the even conserved quantities are non-zero, indicating the breaking of the spatial parity together with the presence of a non-zero current in the solutions. 
Such continuation corresponds to a true phase transition, since the conserved quantities are not smooth functions of the coupling parameter $g$ at $g_{c2}$ \cite{krajenbrink2017exact}.
Indeed, as was noticed numerically in \cite{janas2016dynamical} the conserved quantity $\Delta_h= h_{{\rm opt},B}(+\infty,t)- h_{{\rm opt},B}(-\infty,t)=
h_{\rm opt}(+\infty,t)- h_{\rm opt}(-\infty,t) = \int_\R \rmd x \partial_x Q(x,t)/Q(x,t)$ (where $h_{\rm opt,B}(x,t)$ is the optimal height for the Brownian IC)
can be considered as an order parameter since it is non-zero for $H = \frac{H_B}{2} > H_{c2}$ and vanishes for $H \leq H_{c2}$. Here we conjecture \cite{SM} 
that $\Delta_h$ can be obtained analytically
and is equal to
\begin{equation} \label{OP} 
 \Delta_h =2\log\frac{\kappa_1}{\kappa_0}\mid_{g=g(H=H_B/2)}
\end{equation}
for $g\in (0,1/\sqrt{2e}]$ and  $H=\frac{H_B}{2}\geq H_{c2}$ and $\Delta_h=0$ for $H\leq H_{c2}$. 
Note that \eqref{OP} can be seen as the $n \to 0$ limit of \eqref{cons}
and is not part of the standard ZS conserved quantities \cite{footnoteSpohn}.
The prediction \eqref{OP} is compared to the numerical results of 
\cite{janas2016dynamical} in Fig.~\ref{fig:hxt}.

In this work, we constructed the explicit solution to the weak noise theory of the KPZ equation for the flat and Brownian 
initial conditions, and obtained the exact optimal height and noise fields. The structure of the solution is richer than
in the case of the droplet IC recently solved in \cite{UsWNT2021}. We have shown that the interplay between solitons with different rapidities 
provides a mechanism for obtaining a phase transition in the large deviation.

\begin{acknowledgments}
\paragraph{Acknowledgments.}  We thank B.~Meerson for sharing the data of the numerical solution to the WNT equations in Fig.  \ref{fig:hxt}.  AK acknowledges support from ERC under Consolidator grant number 771536 (NEMO)
and PLD from the ANR grant ANR-17-CE30-0027-01 RaMaTraF.
\end{acknowledgments}

\newpage
\begin{widetext} 

%%%%%%%%%%% Merge with supplemental materials %%%%%%%%%%
%\pagebreak
%%\widetext
%\begin{widetext} 
%\begin{center}
%\textbf{\large Supplemental Materials: Title for main text}
%\end{center}
%%%%%%%%%%% Merge with supplemental materials %%%%%%%%%%
%%%%%%%%%%% Prefix a "S" to all equations, figures, tables and reset the counter %%%%%%%%%%
%\setcounter{equation}{0}
%\setcounter{figure}{0}
%\setcounter{table}{0}
%\setcounter{page}{1}
\makeatletter
\renewcommand{\theequation}{S\arabic{equation}}
\renewcommand{\thefigure}{S\arabic{figure}}
%\renewcommand{\bibnumfmt}[1]{[S#1]}
%\renewcommand{\citenumfont}[1]{S#1}
%%%%%%%%%%% Prefix a "S" to all equations, figures, tables and reset the counter %%%%%%%%%%

\setcounter{section}{0}
\renewcommand{\thesubsection}{S-\Alph{subsection}}

\setcounter{secnumdepth}{2}

\begin{large}
\begin{center}

Supplementary Material for\\  {\it Inverse scattering solution of the weak noise theory of the Kardar-Parisi-Zhang equation with flat and Brownian initial conditions}

\end{center}
\end{large}

We give the principal details of the calculations described in the main text of the Letter. 
We also give additional information about the results displayed in the text.

\section{Relation Brownian-Flat}

\subsection{Previous results for the Brownian IC}

Let us recall the results of \cite{krajenbrink2017exact}, which were obtained by a completely different method making use
of the exact determinantal solution available for the stationary KPZ equation at any finite time. 
There, the following generating function was computed, see Eq (119) in the Supp. Mat.
or Eq. (18) in the limit $\tilde w \to 0^+$, see also discussion around \cite[Formula (7.3.21)]{krajenbrink2019beyond}, together
with its small time large deviation form, for $\tilde{z}>0$
\be \label{PsiB} 
%\int dH_B P(H_B,T) \exp( - \frac{z}{T} e^{H_B + \chi} ) %\overline{ \exp( - \frac{z}{T} e^{H + \chi} )  } 
%\sim 
\int_\R \rmd H_B P(H_B,T) \exp\big( - \frac{2 \sqrt{\tilde{z}}}{\sqrt{T}} e^{H_B/2} \big) 
\sim \exp \big( - \frac{ \Psi_B(\tilde{z})}{\sqrt{T}} \big) 
\ee 
Note that the argument in \eqref{PsiB} is $2 \sqrt{\tilde{z}}$, for technical reasons.
The result for $\Psi_B(\tilde{z})$ obtained in \cite{krajenbrink2017exact} reads
\be \label{eq:brownianExact2017Psi}
\Psi_B(\tilde{z})= \Psi_{B,0}(\tilde{z}) =  - \int_\R \frac{\rmd q}{2 \pi} {\rm Li}_2(- \frac{\tilde{z}}{q^2}  e^{-q^2}) 
\ee 
corresponding to the main branch. One defines the continuation of this function in the two other branches
\bea 
&& \Psi_B(\tilde{z})= \Psi_{B,0}(\tilde{z})  +\Delta_{B,0}(\tilde{z}) \quad , \quad \text{second branch} \\ 
&& \Psi_B(\tilde{z})= \Psi_{B,0}(\tilde{z})  + \frac{\Delta_{B,0}(\tilde{z}) + \Delta_{B,1}(\tilde{z})}{2}  \quad , \quad \text{third branch}
\eea 
where the jump functions are expressed in terms of the Lambert functions $W_0, W_{-1}$ \cite{corless1996lambertw} as  
\bea \label{DeltaBrownien} 
&&   \Delta_{B,0}(\tilde{z})=\frac{4}{3}[-W_0(-\tilde{z})]^{3/2}-4[-W_0(-\tilde{z})]^{1/2}  \\
&&    \Delta_{B,1}(\tilde{z})=\frac{4}{3}[-W_{-1}(-\tilde{z})]^{3/2}-4[-W_{-1}(-\tilde{z})]^{1/2}
\eea

Once the function $\Psi_B(\tilde{z})$ is known the rate function $\Phi_B(H_B)$ is obtained via a Legendre transform, 
which reads explicitly
\bea
\label{PhiResult}
&&\Phi_B(H_B) = 
\begin{cases}
 \displaystyle \max_{\substack{\tilde{z} \in [0,+\infty[}} [\Psi_{B,0}(\tilde{z})  -2\sqrt{\tilde{z}e^{H_B}} ], \; H_B \leq H_{c,B}=0\\
\\
 \displaystyle \max_{\tilde{z} \in [0,e^{-1}]} [\Psi_{B,0}(\tilde{z}) +\Delta_{0,B}(\tilde{z}) +2 \sqrt{ \tilde{z}e^{H_B}} ], \; H_{c,B}\leq H_B \leq H_{c2,B}\\
  \\
 \displaystyle \min_{\tilde{z} \in ]0,e^{-1}]} [\Psi_{B,0}(\tilde{z}) +\frac{\Delta_{0,B}(\tilde{z})+\Delta_{1,B}(\tilde{z})}{2} +2\sqrt{  \tilde{z}e^{H_B}}], \; H_B \geq H_{c2,B} \\
\end{cases}
\eea
with $H_{c2,B}= 2 H_{c2}$, where
\be \label{Hc2} 
H_{c2}= \log \Psi'(z)|_{z=-g_{c2}} = 
\log(\Psi'_0(-\frac{1}{\sqrt{2e}})+\Delta'_0(-\frac{1}{\sqrt{2e}}))   \approx 0.926581 
\ee 
is defined in the Letter. Note that one can understand the change of sign in front of $2\sqrt{\tilde{z} e^{H_B}}$ 
in \eqref{PhiResult} as follows: we first decrease $\tilde{z}$ from $+\infty$ to $0$ and then increase it to $e^{-1}$. 
In the complex $\tilde{z}$-plane, turning around $0$ induces a branch change in the square root function $\sqrt{\tilde{z}}\to -\sqrt{\tilde{z}}$. The change from a maximum to a minimum can be seen from a change of convexity in the argument of the variational problem.\\

\subsection{From the exact solution of the WNT for flat IC to the one for the Brownian IC}

In the paper \cite{Meerson_flatST} the symmetries of the WNT action were studied in the case of the Brownian IC. The authors cleverly noticed that they imply that at time $t_B=1/2$ the KPZ height field 
is flat, i.e. $Q_B(x,1/2)$ is independent of $x$ (where for clarity we denote $t_B \in [0,1]$ the time for the Brownian IC). From this they concluded that
one can deduce the WNT solution for the flat IC if one knows the solution for the Brownian IC. Using our result in \cite{krajenbrink2017exact}, recalled 
in the previous section, they displayed the solution for the flat IC, expected from these symmetries. They obtained the following relation between the
rate functions, which read in our notations
\be 
\Phi(H)= \frac{1}{2 \sqrt{2}} \Phi_B(2 H)   \quad , \quad H < H_{c2} 
\ee 
valid for the main and second branch. In the third branch, there are in fact three solutions to the WNT equations: one is relevant for the flat IC, and 
the two others for the Brownian, as discussed in the text.

In the text we have done the converse: we have obtained directly the solution for the flat IC (which had not been obtained direcly before) denoted
$P(x,t),Q(x,t)$ in the text. We noticed that it can be extended for $t \in [-1,1]$ instead of the original interval $[0,1]$.
From this extension we constructed using \eqref{PQB_flat} the solution $P_B(x,t_B),Q_B(x,t_B)$ (with $t_B=2 t-1$) for the Brownian IC. \\

Let us now give the arguments in support of this construction. The method makes use
of the non-trivial "fluctuation dissipation" symmetry of the dynamical action for the KPZ equation,
and of its implementation on the saddle point equations of the WNT, used in \cite{Meerson_flatST}
(for earlier applications of this symmetry see \cite{canet2011nonperturbative}). We first recall the following general property of the $\{ P,Q \}$ system. Let us define $\tilde Q(x,t)$
and $\tilde P(x,t)$ via the relations
\be \label{eq:symmetry1} 
\tilde Q(x,t)=1/Q(-x,-t)
\ee
and 
\be \label{eq:symmetry2} 
2 g \tilde P(x,t) \tilde Q(x,t) + \partial_x^2 \log \tilde Q(x,t) =
2 g P(-x,-t) Q(-x,-t) + \partial_x^2 \log Q(-x,-t)
\ee
One can show that if $P,Q$ are solutions of \eqref{eq:PQsystem} (with coupling $g$) in some time interval, 
$\tilde P,\tilde Q$ are also solutions of \eqref{eq:PQsystem} (with the same coupling $g$) in the mirror image interval.\\

We now use this symmetry to define an extended solution of the $\{P,Q\}_g$ system \eqref{eq:PQsystem}, $P_F,Q_F$ on the 
interval $t \in [-1,1]$, such that 
\be \label{eq:QF} 
Q_F(x,t)= \begin{cases}  Q(x,t)  , \quad \text{for} \; t \in [0,1] \\ 
\tilde Q(x,t)  , \quad \text{for} \; t \in [-1,0]
\end{cases} 
\ee
and similarly for $P_F$. Let us now define the functions $P_B$ and $Q_B$ for $t \in [0,1]$ as
\be
\begin{split}
    \label{PQB} 
&Q_B(x, t)= e^{\frac{H_B}{2} } Q_F(\sqrt{2} x, 2 t - 1) \\
&P_B(x, t) = \sqrt{2} P_F(\sqrt{2} x, 2 t - 1)
\end{split}
\ee
One can check that $P_B,Q_B$ satisfy the $\{P,Q\}_{g_B}$ system \eqref{eq:PQsystem} with coupling constant $g_B=\sqrt{2} g e^{-H_B/2}$
and $Q_B(0,1)=e^{H_B}$ with $H_B=2 H$. The important point for us now is that if $P,Q$ satisfy the boundary conditions 
for the flat IC 
\be 
Q(x,0)=1 \quad , \quad P(x,1)=\delta(x) 
\ee 
then, $P_B,Q_B$ constructed as above satisfy the boundary conditions for the Brownian IC, which read \cite{janas2016dynamical} 
\begin{itemize}
    \item \textit{(i)} $Q_B(0,0)=1$,
    \item \textit{(ii)} $P_B(x,1)=\delta(x)$,
    \item \textit{(iii)} $g_B P_B(x,0) Q_B(x,0) + \partial_x^2 \log Q_B(x,0) = g_B e^{H_B} \delta(x)$,
\item \textit{(iv)} $Q_B(0,1)=e^{H_B}$. 
\end{itemize}

This can be checked using all the above definitions. For \textit{(i)} one has
\be 
Q_B(0,0)= e^{\frac{H_B}{2} } Q_F(0, - 1) = e^{\frac{H_B}{2} } \tilde Q(0, - 1) = e^{\frac{H_B}{2} }/Q(0,1) = e^{\frac{H_B}{2} - H} = 1 
\ee 
For \textit{(ii)} it is obvious. For \textit{(iii)}, denoting $y=\sqrt{2} x$ and using $g_B=\sqrt{2} g e^{-\frac{H_B}{2} }$
\be
\label{boundaryB2} 
\begin{split} 
 g_B P_B(x,0) Q_B(x,0) + \partial_x^2 \log Q_B(x,0) &= \sqrt{2} g_B e^{\frac{H_B}{2} }  \tilde P(y, - 1)  \tilde Q(y, - 1) 
+ 2 \partial_y^2 \log \tilde Q(y, - 1) \\
& = 2 g \tilde{P}(y, -1) \tilde{Q}(y, -1) 
+ 2 \partial_y^2 \log \tilde Q(y,-1)\\
& = 2 g  P(-y, 1) Q(-y, 1) \\
&= 2 g e^H \delta(y) \\
&=\sqrt{2} g_B e^{H_B}  \delta(\sqrt{2}x)\\
& = g_B e^{H_B} \delta(x)
\end{split}
\ee
where in the third line we have used the symmetry \eqref{eq:symmetry2} and the flat IC. 
For \textit{(iv)} $Q_B(0,1)=e^{H_B/2} Q(0,1)= e^{H_B}$ using $H_B = 2 H$.
Note that \eqref{eq:symmetry1} is continuous at $t=0$ since $Q(x,t=0)=1$. Hence $P_B,Q_B$ constructed as above are the
solution of the WNT for Brownian initial conditions. \\

In the previous paragraph we constructed $P_F(x,t),Q_F(x,t)$ using symmetries. It is not a priori obvious that these functions
should coincide with $P(x,t),Q(x,t)$ extended to the interval $t \in [-1,1]$ as constructed in the text. It turns out 
that this is the case and one has 
\begin{equation} 
\label{eq:QQF} 
 Q(x,t)=Q_F(x,t), \quad P(x,t)=P_F(x,t),   \quad t \in [-1,1]\, .
\end{equation} 
This implies, from \eqref{eq:symmetry1} and \eqref{eq:symmetry2} that the solutions obtained in the text for
$P(x,t)$, $Q(x,t)$ should satisfy, for $t \in [-1,1]$
\be  \label{inverse} 
Q(x,t) Q(-x,-t)=1
\ee 
as well as
\bea
&2g P(x,t) Q(x,t) + \partial_x^2 \log Q(x,t) =
2g P(-x,-t) Q(-x,-t) + \partial_x^2 \log Q(-x,-t) \nonumber
\eea 
These conditions are highly non trivial to check on the analytical form of the solutions provided in the text.
Thus we have performed some numerical checks, e.g. we have checked numerically that the symmetry \eqref{inverse} holds, see below in Section~\ref{app:numerical_evaluation}.
\\

Note that all the above construction is correct for each given branch of solutions. 
For $H = \frac{H_B}{2} \leq H_{c2}$ one thus inserts in \eqref{eq:QF}, \eqref{PQB} the solution $P,Q$ for the flat IC given in the text for the main and
second branch, and one obtains the solution for the Brownian IC for $H_H \leq 2 H_{c2}$.
For $H \geq H_{c2}$
(third branch) there are three simultaneous solutions, as discussed in the text. One of these solutions
(with the choice $\{ \kappa_0,\kappa_0\} $ for the solitonic rapidities)
is even in $x$ and corresponds to the flat IC solution. This solution does not allow to 
obtain the solution for the Brownian IC (it corresponds to a subleading contribution to the 
dynamical action). The two other solutions, 
(with the choice $\{ \kappa_1,\kappa_0 \}$ and $\{\kappa_0,\kappa_1\}$ for the solitonic rapidities)
denoted as $P^\pm,Q^\pm$ in the text, break the $x \to -x$ symmetry and are mirror image of each other.
These are the solution which should be inserted in \eqref{eq:QF}, \eqref{PQB} to obtain the 
solution for the Brownian IC in that regime. Note the symmetries
\eqref{eq:symmetry1} and \eqref{eq:symmetry2} are never broken for any of these solutions, irrespective of
whether $x \to -x$ is broken or not.

\subsection{Rate functions: relations between flat to Brownian}

Let us recall our result in the text for the rate function $\Psi(z)$ for the flat IC in the main branch $z>0$. It reads
\be \label{eq:psi0_z2}
\Psi(z) = \Psi_0(z) := - \int_\R \frac{\rmd q}{4 \pi} {\rm Li}_2(- \frac{z^2}{q^2}  e^{-2q^2}) 
\ee
Comparing with the result for the rate function $\Psi_B(\tilde z)$ for the Brownian initial condition \eqref{eq:brownianExact2017Psi}
in the main branch, we see that the following relation holds
\begin{equation} \label{relation0} 
    \Psi_0(z)=\frac{1}{2\sqrt{2}} \Psi_{B,0}(\tilde z= 2 z^2)
\end{equation}

Let us recall that the rate functions $\Psi_0$ and $\Psi_{B,0}$ are related to the rate functions $\Phi(H)$ and $\Phi_B(H_B)$ 
in the main branch through the Legendre transform
\bea  \label{Legendre3} 
&& \Psi_0(z)=\min_H ( \Phi(H) + z e^H  ) \\
&& \Psi_{B,0}(\tilde z) = \min_{H_B} ( \Phi_B(H_B) + 2 \sqrt{\tilde z e^{H_B}} ) 
\eea
One can easily verify that this is compatible with the relation obtained in \cite{Meerson_flatST} 
\be  
\Phi(H)= \frac{1}{2 \sqrt{2}} \Phi_B(2 H)   
\ee 
This is easily checked inserting $\Phi(H)$ from this relation into the first equation in 
\eqref{Legendre3} and defining $z=\sqrt{\tilde z/2}$.  In fact the relation
\begin{equation}
    \Psi(z)=\frac{1}{2\sqrt{2}} \Psi_{B}(\tilde z= 2 z^2)
\end{equation}
holds for each branch and each solution. As a consequence the jumps are also related. One has 
\be 
\Delta_{0}(z)= \frac{1}{2\sqrt{2}} \Delta_{0,B}(\tilde z= 2 z^2)
\ee 
as can be checked by comparing \eqref{De0} and \eqref{DeltaBrownien}. The same relation holds between $\Delta_1(z)$
and $\Delta_{B,1}(z)$. 
Finally in the third branch the spatially asymmetric solutions discussed in the text associated to 
$\Psi(z)=\Psi_0(z)+ \frac{\Delta_0(z) + \Delta_1(z)}{2}$ correspond to the result in 
the third line of \eqref{PhiResult} for the Brownian initial condition via the same relation. 
\\

{\bf Remark}. In \cite{krajenbrink2017exact} we have obtained the series expansion
\begin{equation}
    \Psi_{B,0}(\tilde z)=\frac{1}{\sqrt{4\pi}}\sum_{n \geqslant 1}(-1)^{n-1}\frac{(4 \tilde z)^{n/2}}{n!}\Gamma\left(\frac{n}{2}\right)\left(\frac{n}{2}\right)^{\frac{n-3}{2}}
\end{equation}
It is useful to note that this provides, using the relation \eqref{relation0} the following series expansion for the rate function
of the flat IC, for $z>0$
\begin{equation}
    \Psi_0(z)=\frac{1}{\sqrt{4\pi}}\sum_{n \geqslant 1}(-1)^{n-1}\frac{(2 z)^{n}}{n!}\Gamma\left(\frac{n}{2}\right)n^{\frac{n-3}{2}}
\end{equation}
\\

{\bf Remark}. We can give an alternative interpretation of the rate function $\Psi(z)$ of the flat IC. Consider now the solution
to the SHE (in rescaled variables) for the
droplet IC considered in \cite{UsWNT2021} and denote it by $Z_\delta(x,t)$. Then one has
\be \label{identity} 
\overline{ \exp\big( - \frac{z}{\sqrt{T}} \int_\R \rmd x \, Z_\delta(x,1) \big)  }  \sim \exp\big( - \frac{\Psi(z)}{\sqrt{T}} \big) 
\ee 
This implies that the PDF of the rate function for the variable $\int_\R \rmd x \, Z_\delta(x,1)$ is the same as $\Phi(H)$
for the flat IC. \\

Indeed, to compute the LHS of \eqref{identity} one performs the same manipulations as in \cite{UsWNT2021} choosing $j(x,t)=-z \delta(t-1)$
in Eq. (6) there. This leads to the $P,Q$ system with boundary conditions $P_\delta(x,1)=1$ and $Q_\delta(x,0)=\delta(x)$. Upon the transformation
\begin{equation}
    Q_\delta(x,t)=P(x,1-t), \quad P_\delta(x,t)=  Q(x,1-t) 
\end{equation}
which leaves invariant the $\{ P,Q \}$ system, one reduces the problem to studying the flat IC and measuring the height field at time $t=1$.  Note that this relation is in fact more general and valid beyond the WNT as an identity in law between the partition function with
flat IC and the integral over space of the partition function with droplet IC (both being the so-called point to line partition sum of directed polymers).

\section{Additional conservation law and order parameter}

In the case considered here where $Q$ does not vanish at infinity, there is an additional non-trivial conservation law which was not discussed in Ref.~\cite{UsWNT2021}. Indeed it is
easy to check, using the equations for the $\{ P,Q \}$ system that 
\be  \label{Qconserved} 
\partial_t \frac{\partial_x Q(x,t)}{Q(x,t)} = \partial_x J_0(x,t) \quad , \quad J_0(x,t)= 2 g P(x,t) Q(x,t) + \frac{\partial_x^2 Q(x,t)}{Q(x,t)}
\ee 
Assuming that $J_0$ vanishes at $x \to \pm \infty$ this implies the conservation law 
\be 
\frac{\rmd}{\rmd t} \int_\R \rmd x \frac{\partial_x Q(x,t)}{Q(x,t)}  = \frac{\rmd}{\rmd t} [ \log Q(+\infty,t) - \log Q(-\infty,t) ] = 0 
\ee 
Note that \eqref{Qconserved} can also be written in terms of the height field and the response (or noise) field (see definitions in 
\cite[Section S-B, Eq.~(S42)-(S43)]{UsWNT2021}) 
\be 
\partial_t \partial_x h(x,t)  = \partial_x \left ( 2 \tilde h(x,t) + \partial_x^2 h(x,t) + (\partial_x h)^2 \right) 
\ee 
which in these variables is simply the time derivative of Eq.~(S42) in \cite{UsWNT2021}.\\

It is interesting to note (although we will not use it here) that a similar conservation equation holds for $P$, i.e.
\be  \label{Pconserved} 
\partial_t \frac{\partial_x P(x,t)}{P(x,t)} =  \partial_x \tilde J_0(x,t) \quad , \quad \tilde J_0(x,t)= - 2 g P(x,t) Q(x,t) - \frac{\partial_x^2 P(x,t)}{P(x,t)}
\ee 
which under similar assumptions implies the conservation of $\log P(+\infty,t) - \log P(-\infty,t)$. \\

Hence the order parameter defined in the text
\be 
\Delta_h = h(+\infty, t) - h(- \infty,t) = \log Q(+\infty,t) - \log Q(-\infty,t) 
\ee 
is time independent. If the solution is even by spatial parity one has $\Delta_h=0$, as is the case for the flat IC and in the
main and second branch for the Brownian IC. If the spatial parity is broken, as in the third branch for the Brownian IC,
it is non-zero. \\

Although we have not attempted to prove it, we believe that this conserved quantity takes a "simple" value in our case. To provide a guess we have examined the value of this quantity in the case of a low-rank soliton. Let us consider as in \cite[Section S-D]{UsWNT2021} the case where
${\cal A}_{xt}$ and ${\cal B}_{xt}$ are rank $n_1$ and $n_2$ operators, respectively, i.e.
${\cal A}_{xt} = \sum_{j=1}^{n_1} q_{\kappa_j} |\kappa_j \rangle \langle \kappa_j|$ and ${\cal B}_{xt} = \sum_{i=1}^{n_2} p_{\mu_i} |\mu_i \rangle  \langle \mu_i|$
and $q_{\kappa_j}=q_{\kappa_j}(x,t) = \tilde q_j e^{- \kappa_j x + \kappa_j^2 t}$ and $p_{\mu_i}=p_{\mu_i}(x,t) = \tilde p_i e^{- \mu_i x - \mu_i^2 t}$
are plane waves. In that case we obtained the formula 
\bea
&& Q(x,t) = \sum_{i,j=1}^{n_1} q_{\kappa_i}  (I + g \sigma \gamma)^{-1}_{ij}   \quad , \quad  \gamma_{ij}= \frac{p_{\mu_i} q_{\kappa_j}}{\mu_i + \kappa_j} \quad , \quad 
\sigma_{ij}= \frac{1}{\kappa_i+\mu_j}
\eea
In the present case we take $n_1=2$ and $n_2=1$ and choose $\kappa_2=0$ and $\tilde q_2 \neq 0$ corresponding to $A_t(x)$ being constant and equal to $\tilde q_2$
as $x \to +\infty$.

\be 
Q(x,t) = \frac{\tilde{q}_1 e^{-\kappa _1 x} \left(g \kappa _1^2
   \tilde{p}_1 \tilde{q}_2 e^{- \mu _1 x}+\mu _1^2
   \left(\kappa _1+\mu _1\right){}^2\right)+\mu _1^2
   \tilde{q}_2 \left(\kappa _1+\mu _1\right){}^2}{g \mu
   _1^2 \tilde{p}_1 \tilde{q}_1 e^{- \kappa _1 x -\mu _1
   x}+\left(\kappa _1+\mu _1\right){}^2 \left(g
   \tilde{p}_1 \tilde{q}_2 e^{- \mu _1 x}+\mu
   _1^2\right)}  |_{\tilde q_1 \to \tilde q_1 e^{\kappa_1^2 t} , \tilde p_1 \to \tilde p_1 e^{- \mu_1^2 t} }
\ee 
It is easy to check that 
\be 
Q(+\infty,t) = \tilde q_2 \quad , \quad Q(-\infty,t) = \tilde q_2 \frac{\kappa_1^2}{\mu_1^2} 
\ee 
hence we find that the order parameter in that case is 
\be \label{Delta} 
\Delta_h = 2 \log \frac{\mu_1}{\kappa_1}
\ee

We believe that this result extends to our case (the asymmetric branches for the Brownian IC)
with $\mu_1 \to \kappa_0$ and $\kappa_1 \to \kappa_1$ where $\kappa_0$ and $\kappa_1$ are defined in the text.
This conjecture is supported by the data in Fig.~\ref{fig:hxt} in the text. \\

{\bf Remark.} Note that in the case of purely solitonic solutions, the standard conserved quantities are equal to
\begin{equation}
    C_n=\frac{\mu_1^n-(-\kappa_1)^n}{n}
\end{equation}
Interestingly, the additional conservation law presented here and Eq.~\eqref{Delta}, although it does not belong to
the standard family of conserved quantities, corresponds to (twice) the limit $\Delta C_n$ for $n \to 0$. \\

{\bf Remark.} In a recent work \cite{SpohnGGE}, a similar-looking additional conservation law, previously missed in the literature, was identified
in a discretized integrable version of the non-linear Schrodinger equation. 
\\

{\bf Remark.} The formula for the order parameter $\Delta_h$ as a function of $H_B$ indicated in the text
\begin{equation} \label{OP2} 
 \Delta_h =2\log\frac{\kappa_1}{\kappa_0}\mid_{g=g(H=H_B/2)}
\end{equation}
is evaluated there explicitly (see Fig.~\ref{fig:hxt}) from the parametric system
\bea \label{0P3} 
&&  \Delta_h =2\log\frac{\sqrt{-W_{-1}(-2 g^2)}}{\sqrt{-W_{0}(-2 g^2)}} \\
&& \Psi'(-g) = e^{H_B/2} 
\eea

\section{More details on the scattering problem}

We give some details on the determination of the scattering amplitudes mentioned in the text.
\\

{\bf Equation for $\bar{\phi}$ at $t=1$. }  Consider the $\partial_x$ equation of the Lax pair for $\bar{\phi}$ at  $t=1$. Using that $P(x,1)=\delta(x)$ it
reads in 
components 
\begin{equation} \label{eq:barphi12} 
\partial_x \bar{\phi}_1=-\I \frac{k}{2} \bar{\phi}_1- g \delta(x)\bar{\phi}_2 \quad , \quad \partial_x \bar{\phi}_2=\I \frac{k}{2} \bar{\phi}_2+ Q(x,1)\bar{\phi}_1\\
\end{equation}
Let us integrate the first equation. Since $\bar{\phi}_1$ vanishes at $-\infty$ it gives
\be \label{eq:phi11} 
\bar{\phi}_1(x,1)=- g e^{- \I \frac{k}{2} x}\Theta(x) \bar{\phi}_2(0,1)
\ee 
Taking the limit $x \to +\infty$, we obtain, from the asymptotics \eqref{eq:plusinfinity} that
\be \label{eq:btilde11} 
\tilde{b}(k,t=1)=- g \bar{\phi}_2(0,1) 
\ee

To determine $\bar{\phi}_2(0,1)$ we can integrate the second equation in \eqref{eq:barphi12}, which gives, using 
\eqref{eq:phi11} and \eqref{eq:btilde11}
\begin{equation}
\begin{cases} \label{eq:phibsolu}
e^{-\I \frac{k}{2} x}\bar{\phi}_2(x,1)=\bar{\phi}_2(0,1) + \tilde{b}(k,1)\int_{0}^x \rmd x' Q(x',1)e^{-\I k x'}, \quad x>0\\
\bar{\phi}_2(x,1)=-e^{\I \frac{k}{2} x}, \quad x<0
\end{cases}
\end{equation}
where in the second equation we have used that $\bar{\phi}_2(x,1)\simeq -e^{\I \frac{k}{2} x}$ for $x \to -\infty$.
Assuming continuity of $\bar{\phi}_2(x,1)$ at $x=0$, this leads to $\bar{\phi}_2(0,1)=-1$ and to
\be  \label{eq:tildeb} 
\tilde{b}(k,t=1)= g \quad   \Rightarrow \quad \tilde b(k)= g e^{-k^2} 
\ee 
since we recall that $\tilde b(k,t)=\tilde b(k) e^{k^2 t}$.\\

Taking the $x\to +\infty$ limit of \eqref{eq:phibsolu} and adding and substracting $c$ we see that it is compatible
with the asymptotics \eqref{eq:plusinfinity} and gives in addition a relation between $\tilde a(k)$ and $Q(x,1)$
\bea \label{eq:tildea} 
&& \tilde a(k) = \tilde{a}(k,1) = 
%\tilde a(k) = 1- g \lim_{x \to + \infty} \left( \int_{0}^{+\infty} \rmd x' Q(x',1)e^{- \I k x'} - \frac{c}{- \I k} e^{-\I k x}\right)  \\
 1- g \lim_{x \to + \infty} \left( \int_{0}^{x} \rmd x' (Q(x',1)-c)e^{- \I k x'} - \frac{c}{- \I k} \right)  \\
&& =  1- g  \int_{0}^{+\infty} \rmd  x' (Q(x',1)-c)e^{- \I k x'} + g \frac{c}{- \I k} 
\eea
\\

{\bf Equation for $\phi$ at $t=1$.}  Consider the $\partial_x$ equation of the Lax pair for $\phi$ at  $t=1$. Using that
using that $P(x,1)=\delta(x)$ it reads in components
\begin{equation} \label{eq:barphi12n} 
\partial_x {\phi}_1=-\I \frac{k}{2} {\phi}_1- g \delta(x) {\phi}_2 \quad , \quad \partial_x {\phi}_2=\I \frac{k}{2} {\phi}_2+ Q(x,1) {\phi}_1\\
\end{equation}
 which can be rewritten as 
\begin{equation} \label{2eq} 
\begin{split}
[e^{\I \frac{k}{2} x}\phi_1(x,1)]'&=- g \delta(x) \phi_2(x,1)e^{\I \frac{k}{2} x}, \qquad [e^{- \I \frac{k}{2} x}\phi_2(x,1)]'=Q(x,1)\phi_1(x,1)e^{- \I \frac{k}{2} x}
\end{split}
\end{equation}

Integrating these two equations, and using the asymptotics \eqref{eq:plusinfinity} at $x \to +\infty$ 
and $\phi_1(x,1) \to e^{- \I k x/2}$ and $\phi_2(x,1) \to \frac{c}{- \I k}  e^{- \I k x/2}$ at $x \to -\infty$, we obtain
\begin{equation}
\begin{split} \label{solusolu} 
\phi_1(x,1)&=e^{-\I  \frac{k}{2} x}(\Theta(-x)+a(k)\Theta(x)), \quad a(k)-1=- g \phi_2(0,1)\\
\phi_2(x,1)&=e^{\I \frac{k}{2} x } \lim_{X \to - \infty} 
\left( \int_{X}^x \rmd  x' Q(x',1)e^{- \I k x'}(\Theta(-x')+a(k)\Theta(x')) + e^{- \I k X} \frac{c}{- \I k} 
\right) 
\end{split}
\end{equation}
where we used that $a(k,t)=a(k)$. The last equation can be rewritten as
\be
 \phi_2(x,1)=e^{\I \frac{k}{2} x } \lim_{X \to - \infty} 
\left( \int_{X}^x \rmd  x' Q(x',1)e^{- \I k x'}(\Theta(-x')+a(k)\Theta(x')) 
- \int_{X}^x \rmd  x' e^{- \I k x'} c + e^{- \I k x} \frac{c}{- \I k} 
\right) 
\ee

Setting $x=0$ we obtain a relation between $\tilde a(k)$ and $Q(x,1)$
\begin{equation} \label{eq:a} 
a(k) = 1 - g \phi_2(0,1)= 1- g \int_{-\infty}^0 \rmd  x' (Q(x',1)-c) e^{- \I k x'} - g \frac{c}{- \I k} 
\end{equation}

Note that integrating the second equation in \eqref{2eq} for $\phi_2(x,1)e^{-\I \frac{k}{2} x}$ between 0 and $+\infty$ and using the asymptotics \eqref{eq:plusinfinity} 
leads to an expression for $b(k)$, however this expression is equivalent to the one obtained from the relation
$a(k) \tilde a(k) + b(k) \tilde b(k)=1$ obtained from the Wronskian (see the main text) together with the above results for
$\tilde b(k),\tilde a(k), a(k)$. \\

From the above results we see that if $Q(x,1)$ is even one has $\tilde{a}(k)=a(-k)=a^*(k)$ (for real $k$).
From the Wronskian relation and \eqref{eq:tildeb} one thus gets $b(k) g e^{-k^2} = 1 - a(k) a(-k) = 1 - |a(k)|^2$, hence
$b(k)$ is real and even in $k$. Alternatively one sees that $|a(k)|$ is fixed by $b(k)$ so one can write
\begin{equation}
a(k)=e^{- \I \varphi(k)} \sqrt{1-g b(k) e^{-k^2}}
%\qquad \tilde{a}(z)=e^{- \I \varphi_\alpha(-2z)} \sqrt{1-\sqrt{\alpha} b(z) e^{-4z^2}}
\end{equation}
where $\varphi(k)$ is a real and odd function $\varphi(k)=-\varphi(-k)$, as discussed in the text.
\\

It is important to note that the analysis of the scattering equation was performed here assuming that the parity is not broken,
which holds for the flat IC. \\

{\bf Remark}. {\it Small $k$ behavior}. Since we expect that $Q(x,1)$ is smooth and decays fast towards $c$ as $x \to \pm \infty$ we 
can extract from the relations obtained above the behavior of the scattering amplitudes as $k \to 0$
\be 
a(k) \simeq g \frac{c}{\I k}  \quad , \quad \tilde a(k) \simeq g \frac{c}{- \I k} 
\ee 
which implies
\be 
b(k) = \frac1{\tilde b(k)} (1 - a(k) \tilde a(k))   \simeq - g c^2 \frac{1}{k^2} 
\ee
which is consistent with \eqref{b} in the text. 
The integrands in the functions $A_t(x)$ and $B_t(x)$ in \eqref{AGeneral}, \eqref{BGeneral}, i.e.
the reflection amplitudes $r(k)$ and $\tilde r(k)$ thus behave respectively for small $k$ as
\bea
r(k) = b(k)/a(k) \simeq - \tilde a(k)/g \simeq \frac{c}{\I k} \quad , \quad \tilde r(k) = \tilde b(k)/(g \tilde a(k)) \simeq \frac{-\I k}{c g}.
\eea
\\

{\bf Remark}. 
{\it Schrödinger equation}. It is interesting to note that the $\partial_x$ equation of the Lax pair can always be written as a Schrödinger equation, albeit
with a complex potential in the general case. One has
\begin{equation} 
\partial_x {\phi}_1(x)=-\I \frac{k}{2} {\phi}_1(x)- g P(x) {\phi}_2(x) \quad , \quad 
\partial_x {\phi}_2(x)=\I \frac{k}{2} {\phi}_2(x)+ Q(x) {\phi}_1(x)\\
\end{equation}
where here $Q(x)=Q(x,t)$, $P(x)=P(x,t)$ and $t$ can be arbitrary and fixed, so we suppress the time variable.
One can eliminate $\phi_1$ and one obtains that $\phi_2$ satisfies
\be 
\phi_2''(x)  -\frac{\phi_2'(x) Q'(x)}{Q(x)}+\left(g P(x)
   Q(x)+\frac{k^2}{4}+\frac{\I k Q'(x)}{2 Q(x)}\right) \phi_2(x) = 0
\ee 

The first derivative term can be eliminated by writing
\bea 
\phi_2(x)  = \sqrt{Q(x)} f_2(x) 
\eea 
where now $f_2(x)$ satisfies a Schrödinger equation
\be 
f_2''(x)+\frac{1}{4} f_2(x) \left(4 g P(x) Q(x)+k^2+\frac{2 \left(Q''(x)+\I k
   Q'(x)\right)}{Q(x)}-\frac{3 Q'(x)^2}{Q(x)^2}\right) = 0 
\ee 
In the general case the potential is complex, and the problem is non-Hermitian. However,
for the flat IC, $Q(x)=c$, it simplifies and one obtaines the simple result given in the text.

\section{Numerical evaluations} \label{app:numerical_evaluation}

In this section we present some additional numerical evaluations which support the results presented in the text. 

\subsection{Functions $\varphi$, $A_t$ and $B_t$} 

First we have plotted in Fig. \ref{fig:varphik} the function $\varphi(k)$ defined in \eqref{soluphase} as a function of $k$. 
It clearly shows that it has a discontinuity at $k=0$ with $\varphi(0^\pm)=\mp \frac{\pi}{2}$ as stated in the text.
\begin{figure}[ht!]
    \centering
    \includegraphics[scale=0.55]{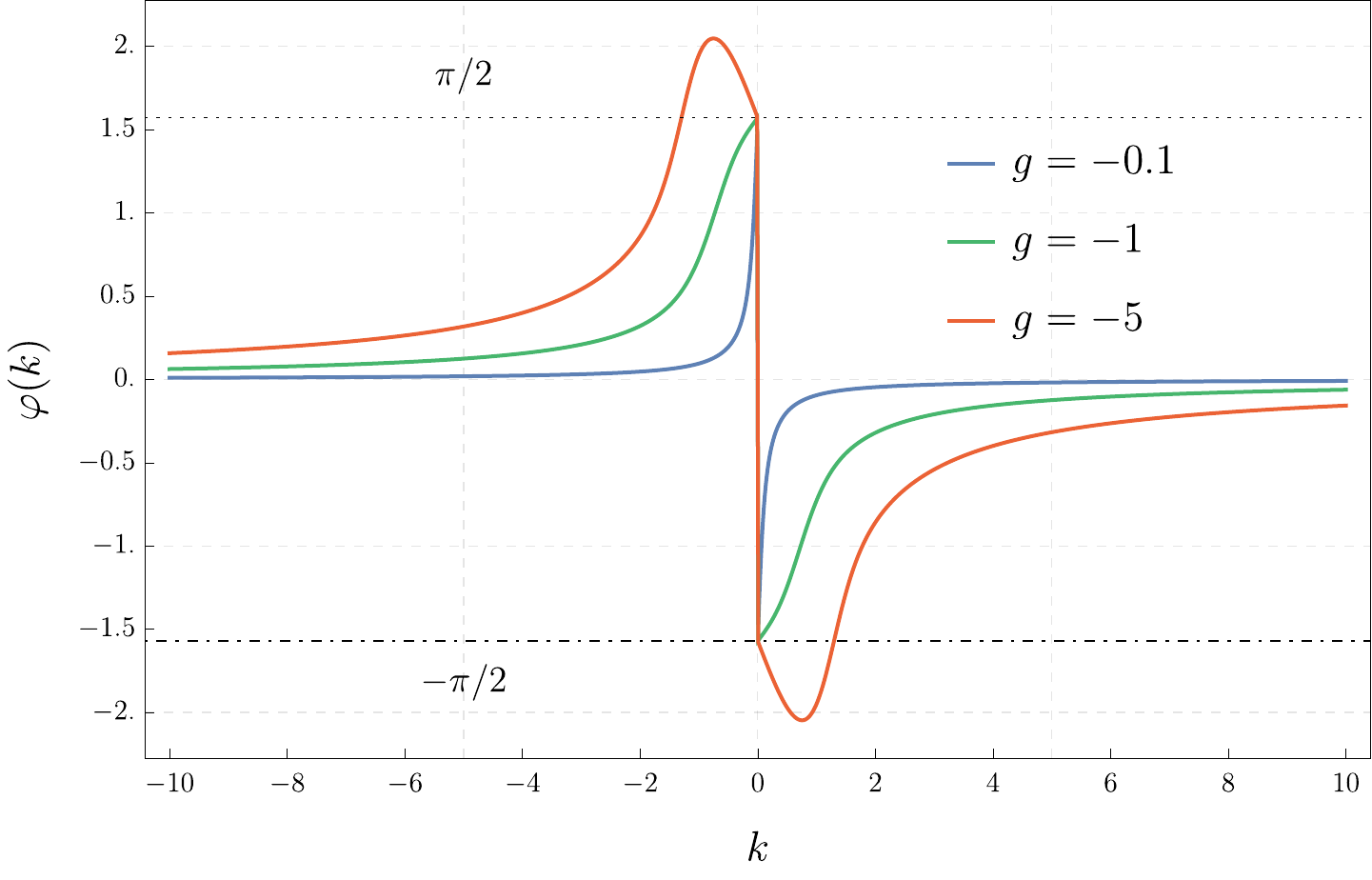}
    \caption{The phase $\varphi(k)$ defined in \eqref{soluphase} plotted versus $k$ for various values of $g$.}
    \label{fig:varphik}
\end{figure}

Next we have plotted in Fig.~\ref{fig:A0} the function $A_t(x)$ for several values of positive time $t$ and $g$
corresponding to the main branch \eqref{AGeneral} and to the second branch \eqref{ABGeneralContinued},
as well as at the critical point $g=g_{c2}$. We recall that the function $g(H)$ is plotted in Fig.~\ref{fig:hxt} in the text. 
\begin{figure}[ht!]
    \centering
    \includegraphics[scale=0.65]{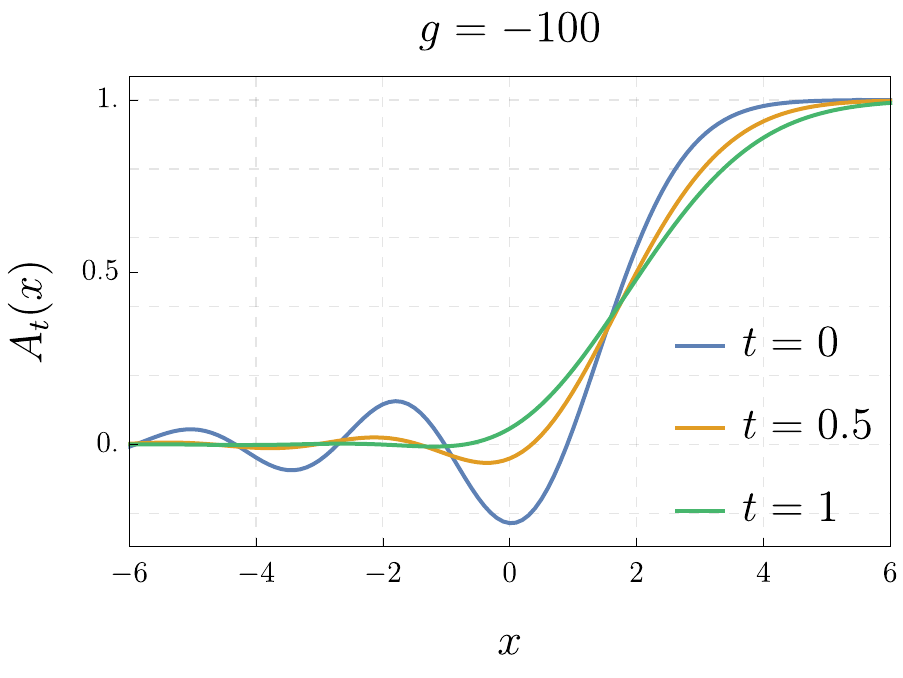}
    \includegraphics[scale=0.65]{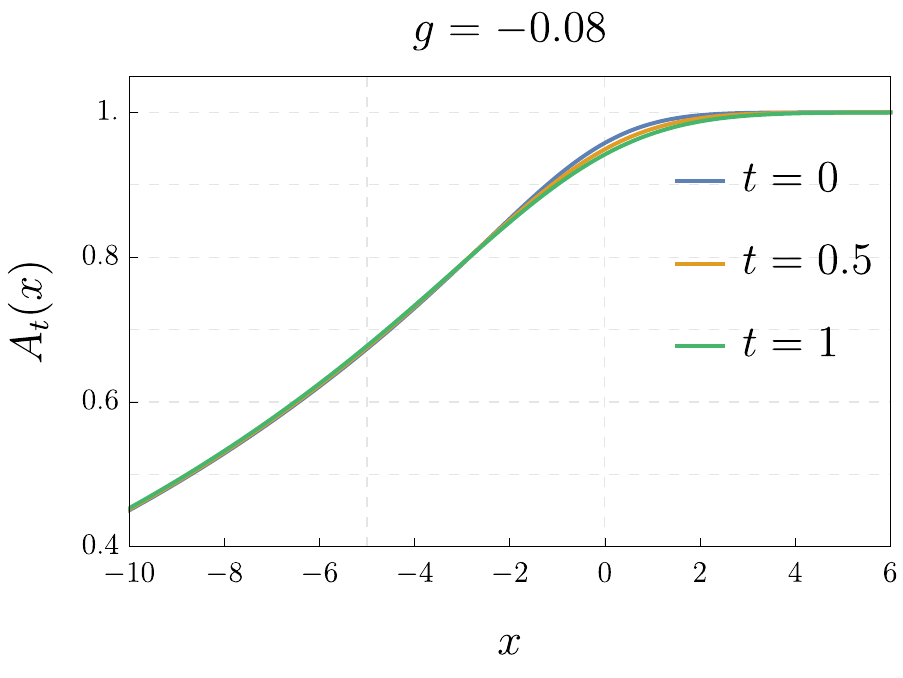}\\
    \includegraphics[scale=0.65]{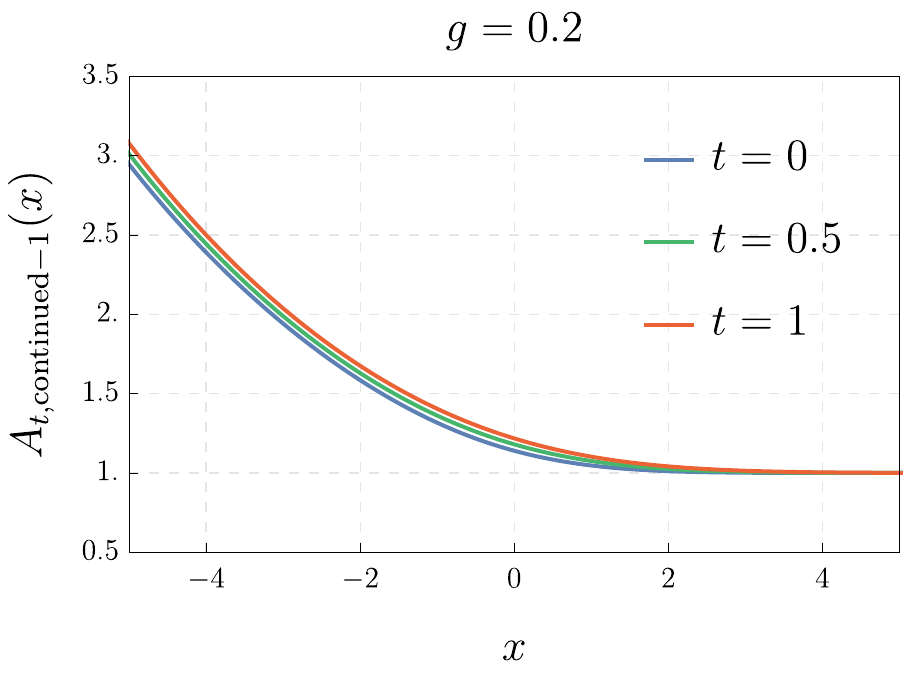}
    \includegraphics[scale=0.65]{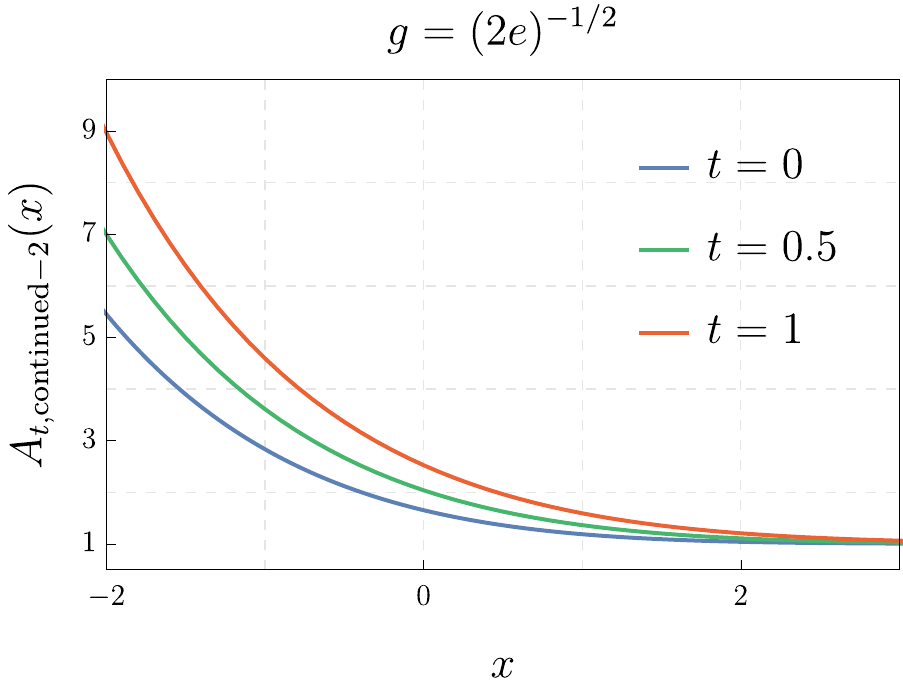}
    \caption{Plot of the function $A_t(x)$ for various positive times $t$, coupling constants $g$ for the main and second branch.}
    \label{fig:A0}
\end{figure}

In Fig.~\ref{fig:B0} we have plotted the function $B_t(x)$ for several values of positive time $t$ and $g$
corresponding to the main branch \eqref{BGeneral} and second branch \ref{ABGeneralContinued}, 
as well as at the critical point $g=g_{c2}$.\\

\begin{figure}[ht!]
    \centering
    \includegraphics[scale=0.65]{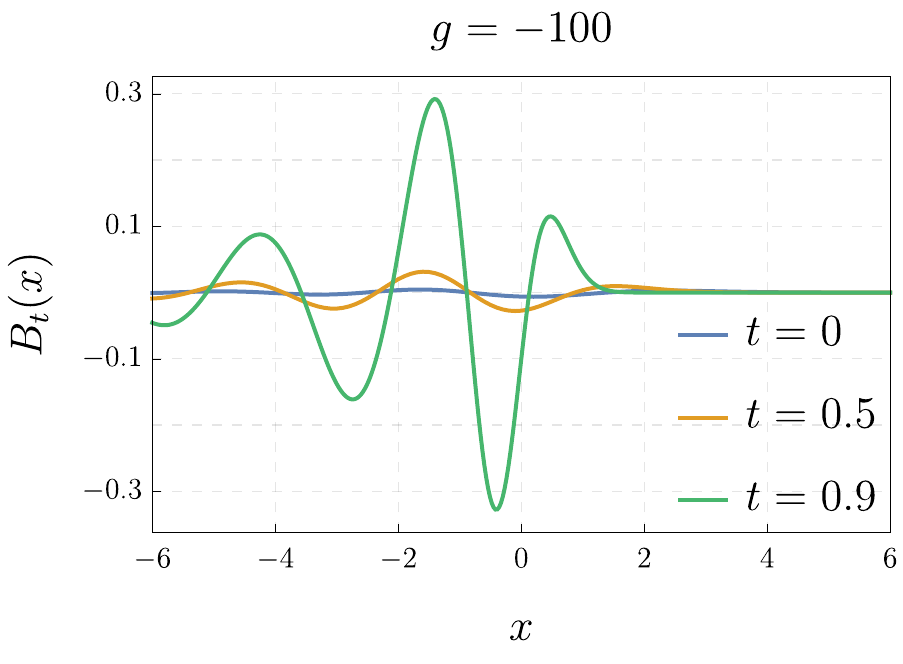}
    \includegraphics[scale=0.65]{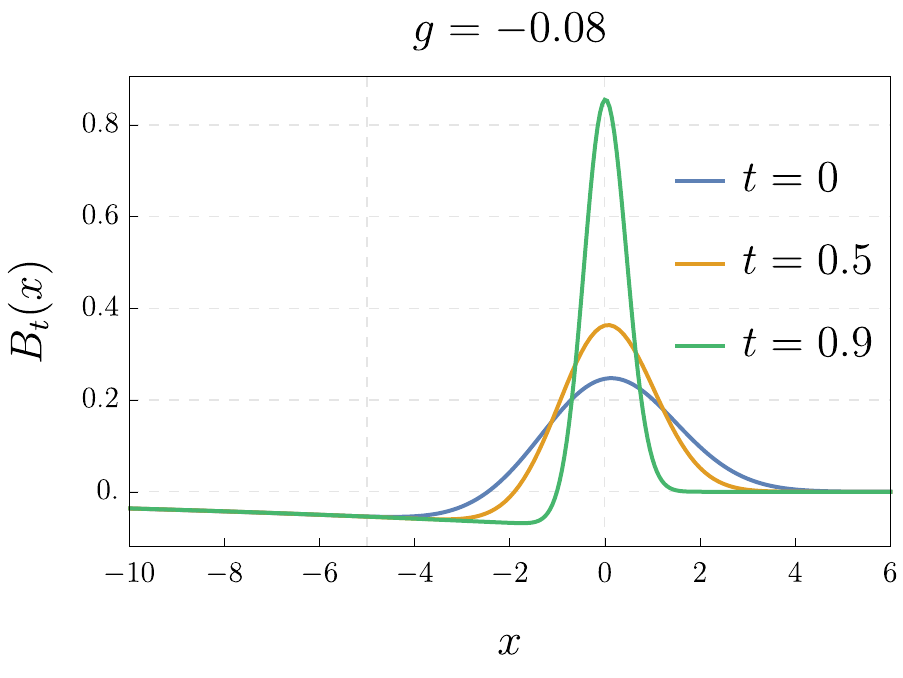}\\
    \includegraphics[scale=0.65]{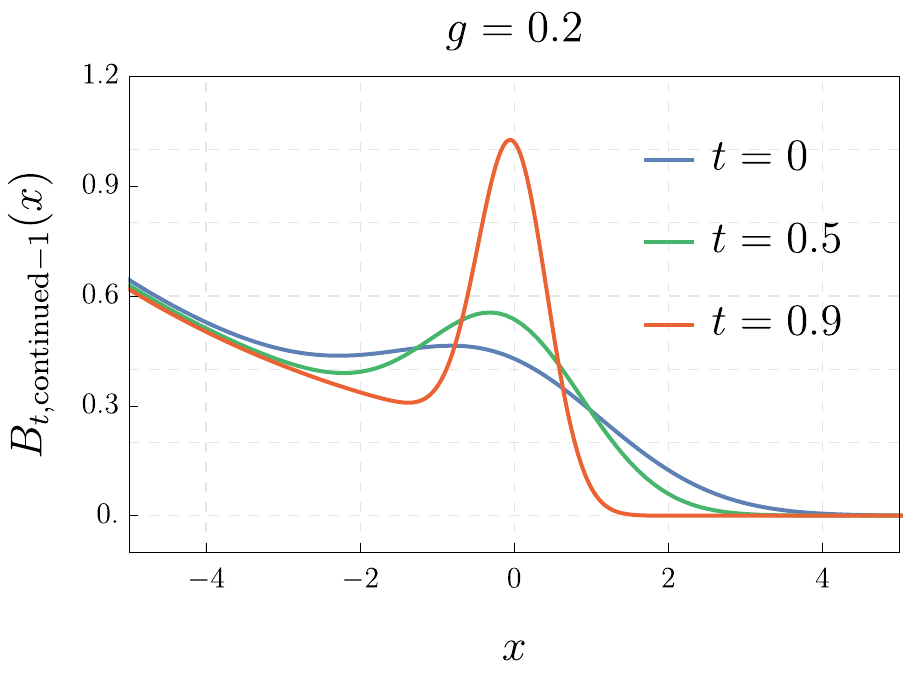}
    \includegraphics[scale=0.65]{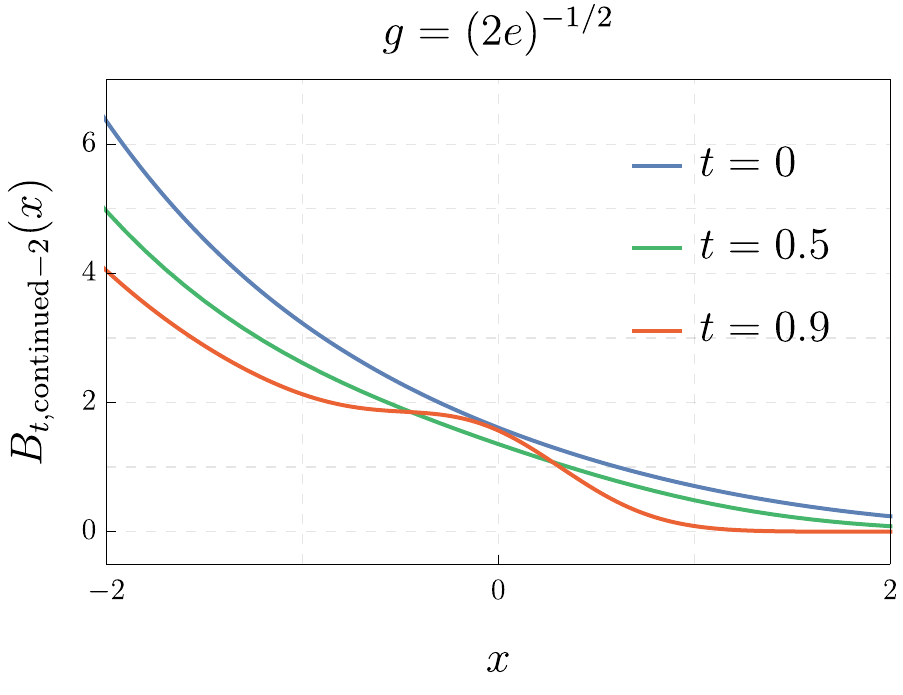}
    \caption{Plot of the function $B_t(x)$ for various positive times $t$, coupling constants $g$ for the main and second branch} 
    \label{fig:B0}
\end{figure}

We have also plotted these functions for negative times (as is of interest for the Brownian IC, see text) for the same values of $g$ for
the main and second branch. 
These are shown in Figs.~\ref{fig:A0_negTime} and \ref{fig:B0_negTime}. Note the relation (see text) 
$A_t''(x)=g B_{-t}(x)$ valid in all the symmetric branches.

\begin{figure}[ht!]
    \centering
    \includegraphics[scale=0.65]{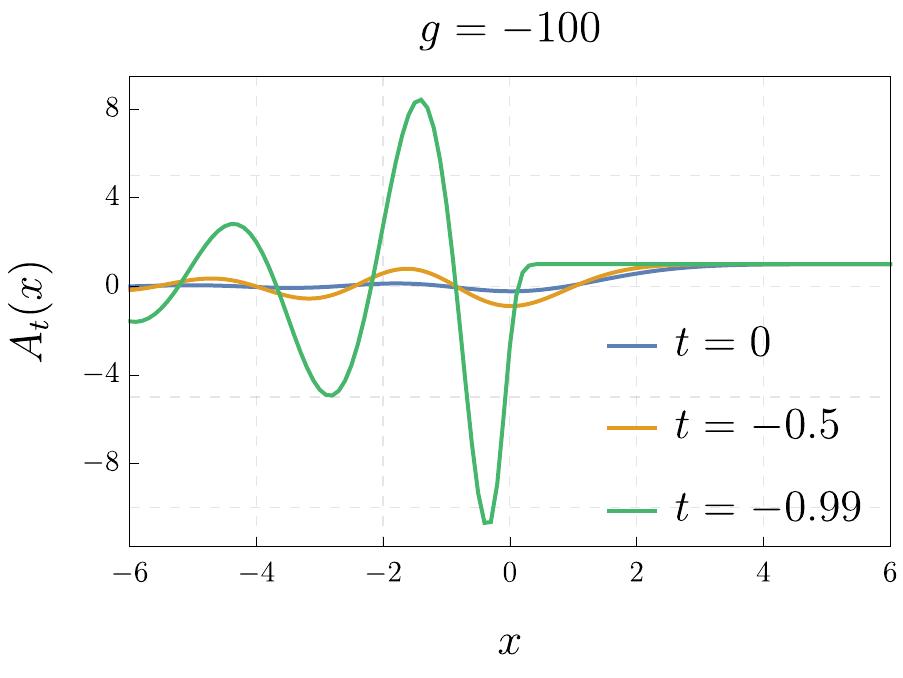}
    \includegraphics[scale=0.65]{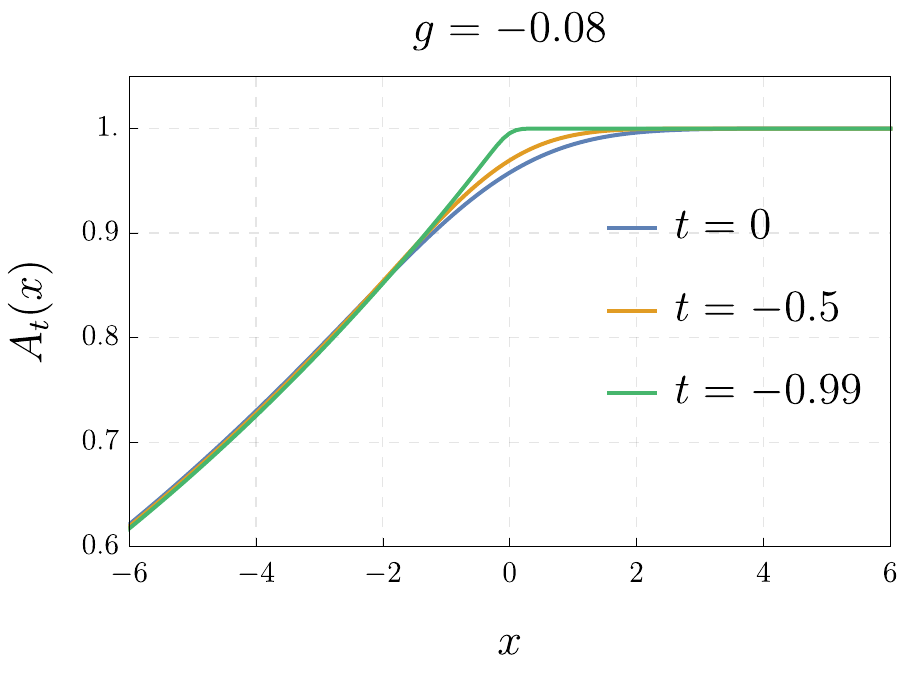}\\
    \includegraphics[scale=0.65]{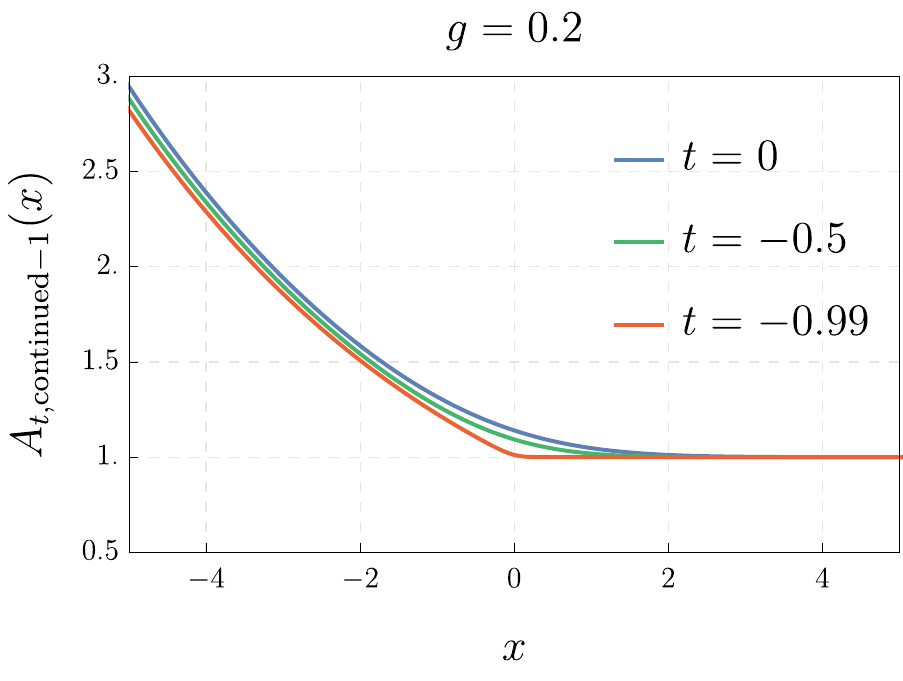}
    \includegraphics[scale=0.65]{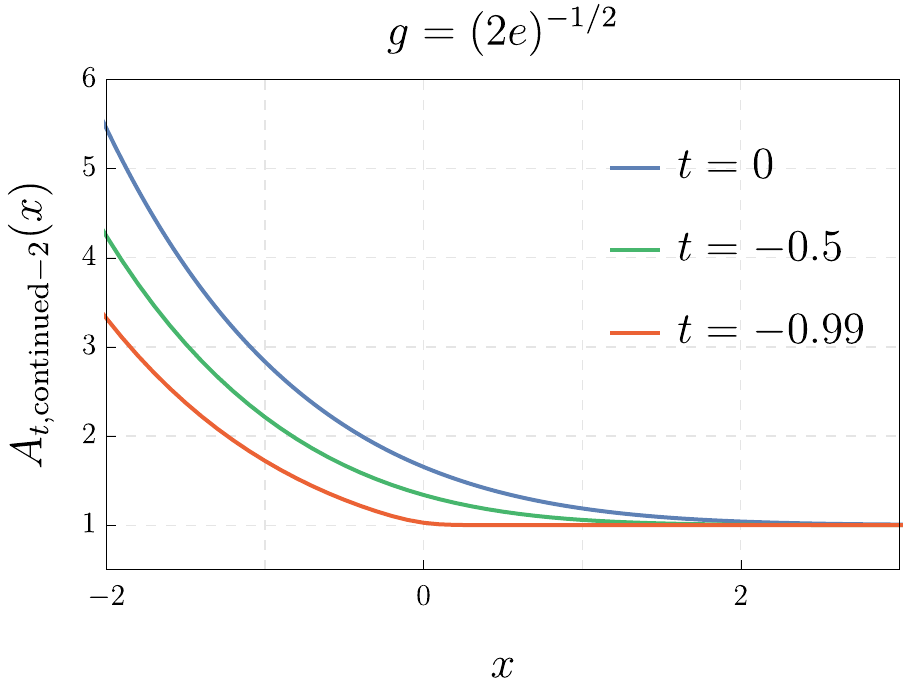}
    \caption{Plot of the function $A_t(x)$ for various positive times $t$, coupling constants $g$ for the main and second branch.}
    \label{fig:A0_negTime}
\end{figure}

\begin{figure}[ht!]
    \centering
    \includegraphics[scale=0.65]{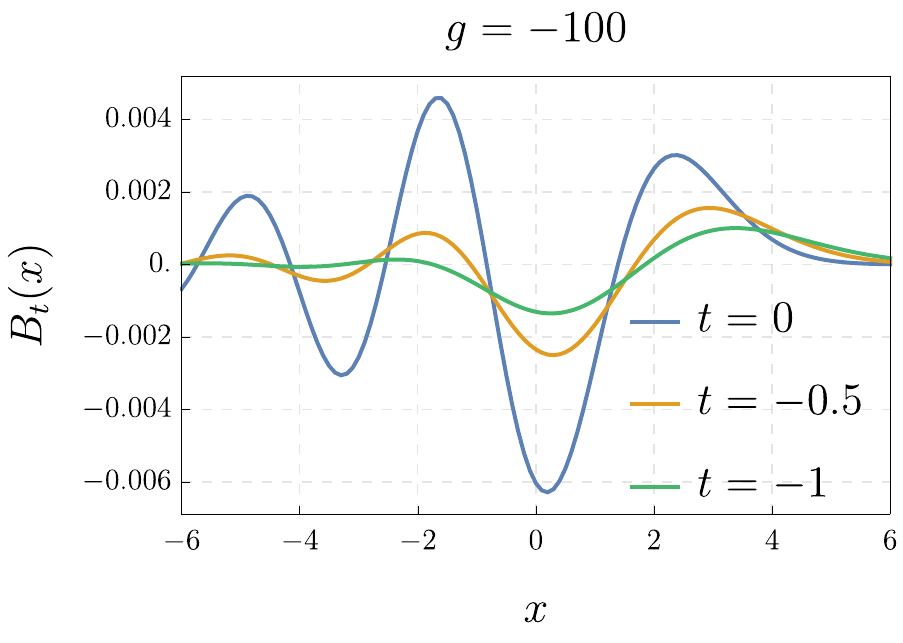}
    \includegraphics[scale=0.65]{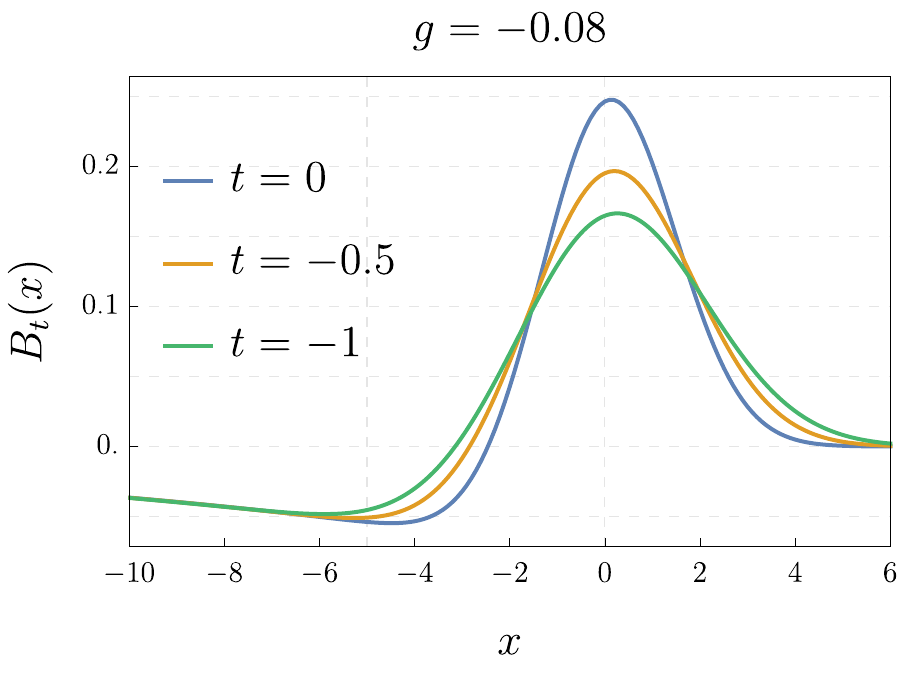}\\
    \includegraphics[scale=0.65]{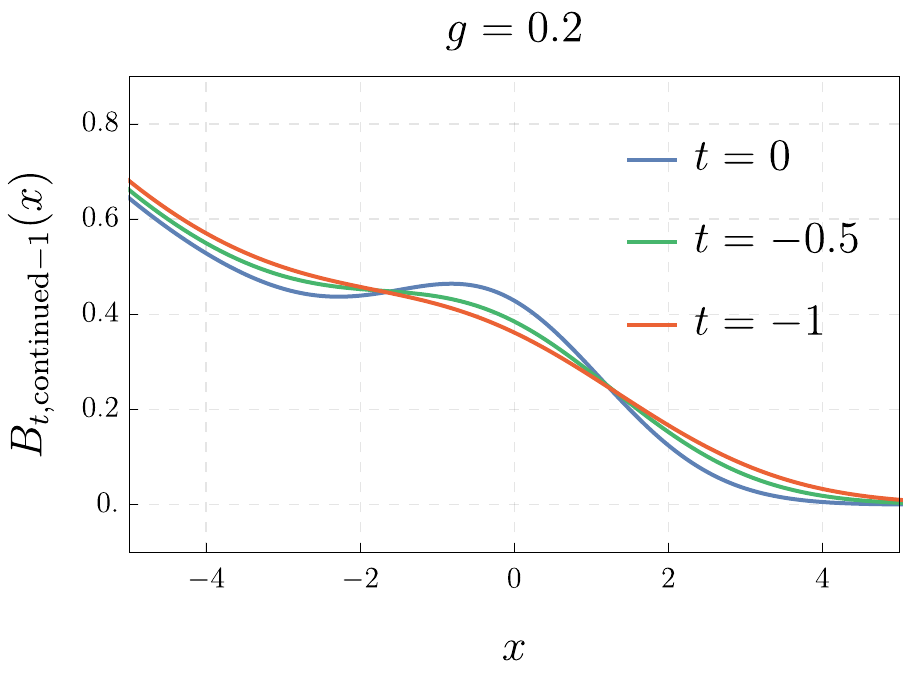}
    \includegraphics[scale=0.65]{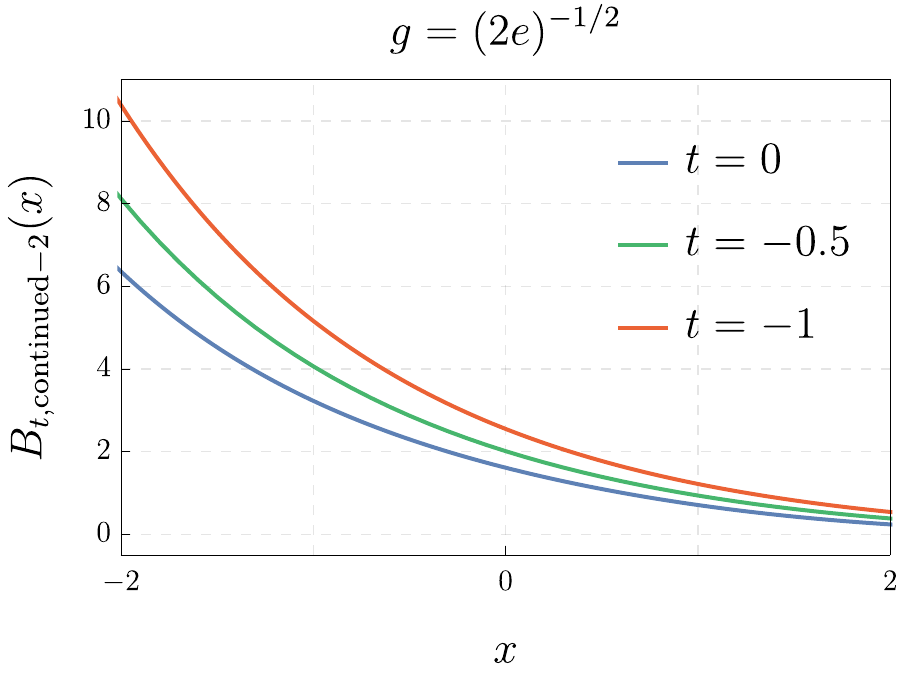}
    \caption{Plot of the function $A_t(x)$ for various positive times $t$, coupling constants $g$ for the main and second branch.}
    \label{fig:B0_negTime}
\end{figure}

\subsection{Optimal height and noise, evaluation of $P,Q$} 

From the above exact solutions for $A_t(x)$ and $B_t(x)$ we obtain the solutions to 
the $\{P,Q\}_g$ system through the Fredholm operator inversion formula \eqref{soluQP} for various
values of $H$ and $g$. We use the numerical method developed in \cite[Section S-L]{UsWNT2021}.\\

For the solution for the flat IC, we have performed several numerical checks of some highly non-trivial consequences of the formulae,
which validate our conjecture:
\begin{itemize}
    \item \textit{(i)} the functions $P,Q$ are even in $x$,
    \item \textit{(ii)}
$Q(x,t=1)=A_1(|x|)$,
\item \textit{(iii)} $Q(0,t=1)=e^H=\Psi'(-g)$,
\item \textit{(iv)} $Q(\pm \infty,t)=1$.
\end{itemize}
We found them to hold in all three branches in the case of the flat IC.  The results for the optimal height $h_{\rm opt}(x,t)=\log Q(x,t)$ are plotted in
Fig.~\ref{fig:hxt}.  Concerning the extension of the flat IC solution to negative times, of interest for the Brownian initial condition, we have also performed a numerical 
check of the symmetry \eqref{inverse} in the main branch, the second branch, and the symmetric third branch.

\section{The Lambert W function} \label{app:lambert}
We introduce the Lambert $W$ function \cite{corless1996lambertw} which we use extensively throughout this work. Consider the function defined on $\mathbb{C}$ by $f(z)=ze^z$, the $W$ function is composed of all inverse branches of $f$ so that $W(z e^z)=z$. It does have two real branches, $W_0$ and $W_{-1}$ defined respectively on $[-e^{-1},+\infty[$ and $[-e^{-1},0[$. On their respective domains, $W_0$ is strictly increasing and $W_{-1}$ is strictly decreasing. By differentiation
 of $W(z) e^{W(z)}=z$, one obtains a differential equation valid for all branches of $W(z)$
\begin{equation} \label{derW} 
\frac{\rmd W}{\rmd z}(z)=\frac{W(z)}{z(1+W(z))}
\end{equation}
Concerning their asymptotics, $W_0$ behaves logarithmically for large argument $W_0(z)\simeq_{z\to +\infty} \log(z)-\log \log (z)$ and is linear for small argument  $W_0(z) =_{z \to 0} z-z^2+\mathcal{O}(z^3)$. $W_{-1}$ behaves logarithmically for small argument $W_{-1}(z)\simeq_{z \to 0^-} \log(-z)-\log(-\log(-z))$. Both branches join smoothly at the point $z=-e^{-1}$ and have the value $W(-e^{-1})=-1$. These remarks are summarized on Fig. \ref{fig:Lambert}. More details on the
other branches, $W_k$ for integer $k$, can be found in \cite{corless1996lambertw}.

\begin{figure}[ht!] 
\begin{center}
\includegraphics[width = 0.6\linewidth]{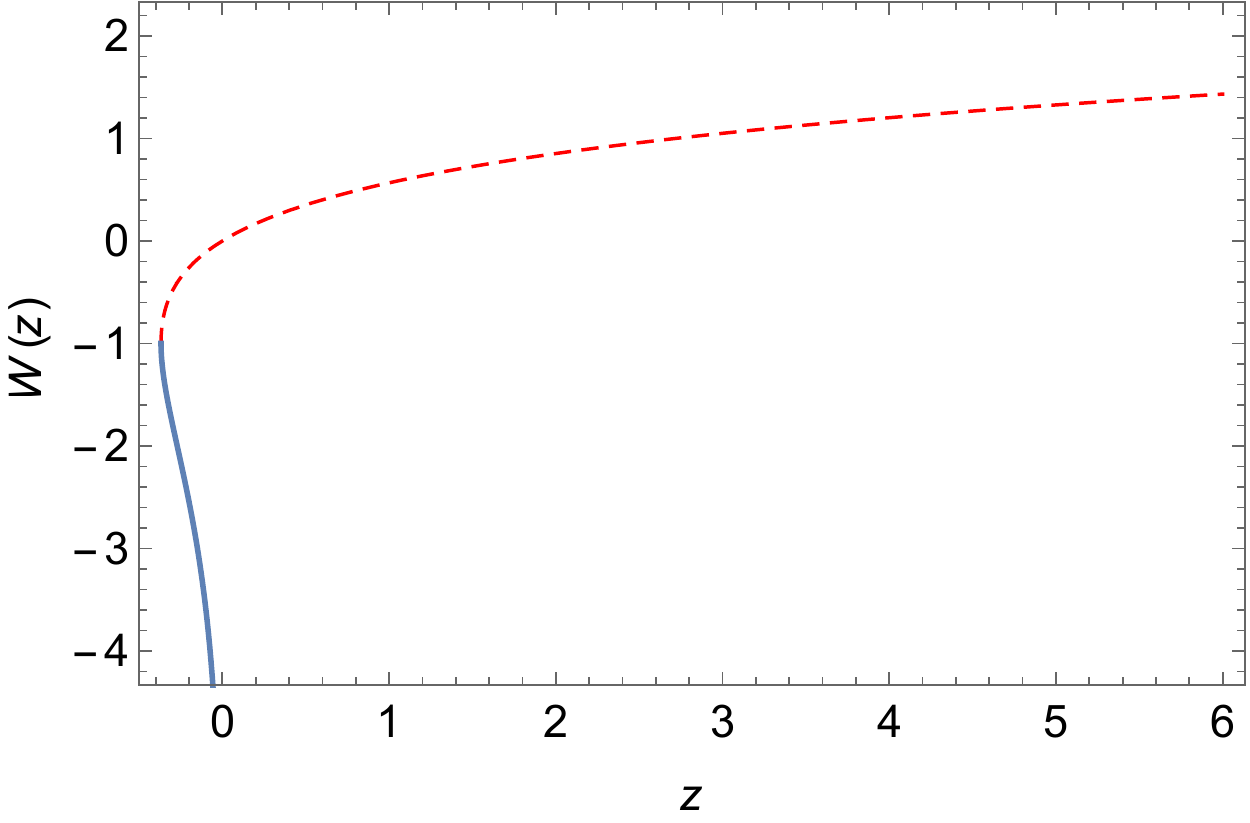}
\caption{The Lambert function $W$. The dashed red line corresponds to the branch $W_0$ whereas the blue line corresponds to the branch $W_{-1}$. }
\label{fig:Lambert}
\end{center}
\end{figure}

\end{widetext}


\begin{thebibliography}{100}
\bibitem{footnoteScalingXT} In this work, we use the same notation $h$ for the height fields $h(y,\tau)$ and $h(x,t)$ and it remains clear that $x,t$ are rescaled space and time variables.
\bibitem{footnoteSpohn} Note that a similar logarithmic conserved quantity has appeared very recently in \cite{SpohnGGE} in the context of an integrable discretized nonlinear Schrodinger equation.
\end{thebibliography}
\end{document}